\documentclass[10pt,twocolumn]{article} 

\usepackage{cite}

\usepackage{amsmath,amssymb,amsfonts}
\usepackage{algorithmic}
\usepackage{graphicx}
\usepackage{textcomp}
\usepackage{xcolor}

\usepackage{caption}
\usepackage{subcaption}
\usepackage{multirow}

\usepackage[ruled,linesnumbered,vlined]{algorithm2e}
\usepackage{upgreek,float}
\usepackage{mathtools}

\title{Security Concerns on Machine Learning Solutions for 6G Networks in mmWave Beam Prediction}

\author{Ferhat Ozgur Catak$^1$ \and Evren Catak$^2$ \and Murat Kuzlu$^3$ \and Umit Cali$^4$ \and Devrim Unal$^5$}

\date{
$^1$ Simula Research Laboratory, Fornebu, Norway, ozgur@simula.no \\
$^2$ Norwegian University of Science and Technology,  Gjøvik, Norway, evren.catak@ieee.org \\
$^3$ Old Dominion University, VA, USA, mkuzlu@odu.edu \\
$^4$ Norwegian University of Science and Technology, Trondheim, Norway, umit.cali@ntnu.no
$^5$ KINDI Center for Computing Research, College of Engineering, Qatar University, Doha, Qatar, dunal@qu.edu.qa
}

\begin{document}

\maketitle

\begin{abstract}
6G --sixth generation-- is the latest cellular technology currently under development for wireless communication systems. In recent years, machine learning algorithms have been applied widely in various fields, such as healthcare, transportation, energy, autonomous car, and many more. Those algorithms have been also using in communication technologies to improve the system performance in terms of frequency spectrum usage, latency, and security. With the rapid developments of machine learning techniques, especially deep learning, it is critical to consider the security concern when applying the algorithms. While machine learning algorithms offer significant advantages for 6G networks, security concerns on Artificial Intelligent (AI) models are typically ignored by the scientific community so far. However, security is also a vital part of AI algorithms because attackers can poison the AI model itself. This paper proposes a mitigation method for adversarial attacks against proposed 6G machine learning models for the millimeter-wave (mmWave) beam prediction using adversarial learning. The main idea behind generating adversarial attacks against machine learning models is to produce faulty results by manipulating trained deep learning models for 6G applications for mmWave beam prediction. We also present a proposed adversarial learning mitigation method's performance for 6G security in mmWave beam prediction application a fast gradient sign method attack. Our results shows that the mean square errors (i.e. the prediction accuracy) of the defended model under attack are very close to the undefended model without attack.
\end{abstract}

\noindent \textbf{keywords: }Machine learning, AI, millimeter-wave, beamforming, adversarial machine learning, 6G, deep learning

\section{Introduction}
Cellular networking has been the most popular wireless communication technology in the last three decades (1G-2G in the early 1990s, 3G in the early 2000s, 4G in 2010s, 5G in 2020s), which can support high data rate with long distance for voice and data. Data transmission speed and the number of users have increased sharply, with recent versions such as 4G (WiMAX and LTE), 4.5G (LTE Advanced Pro), 5G, and 6G. Cellular systems typically operate over land areas, called cells, served by fixed-based transceiver stations, i.e., base stations (BSs), in various frequency bands from 850 Mhz to 95 GHz \cite{6515173}. Latest cellular technologies (4G/5G/6G) support higher data rates, i.e., approximately 33.88 Mbps, 1.100 Mbps, and 1 Tbps, respectively, and low latency, i.e., milliseconds. However, they are still suffering congestion and reduced network performance due to sharing the frequency spectrum with other mobile users.

Introducing the 5G with super-fast data speeds is a breakthrough and presents a significant transformation in mobile networking and data communication. It offers a data transmission speed of 20 times faster than the 4G networks and delivers less than a millisecond data latency \cite{9040431}, \cite{ccatak2016waveform}, \cite{CATAK2017184}. The main difference of 5G is to use a new technology called massive multiple-input multiple-output (MIMO) and using multiple targeted beams to spotlight \cite{6736750}, \cite{6815892}.

Authors in \cite{6848613} investigate several MIMO architectures and MIMO and beamforming solutions for 5G Technology. According to the results, the precise antenna array calibration with large-scale antenna arrays for multi-user-MIMO (MU-MIMO) is needed. MIMO can also enable more devices to be used within the same geographic area, i.e., 4.000 devices per square kilometers for 4G, while around one million for 5G \cite{6736746}. 6G is the last version of this series, which follows up on 4G and 5G. It promises mobile data speeds of 100 times faster with lower latency than the 5G network, i.e., approximately 1 Tbps and 1 ms, respectively. Although the primary use cases of 6G are still under definition, it is clear that 6G will be used in the connectivity in cars, drones, mobile devices, IoT devices, homes, industries, and many more. One of the primary and fundamental differences of 6G technologies is the use of artificial intelligence (AI) and edge computing to make data communication networks more sophisticated \cite{8922617}, \cite{8395149}. Using the benefits of the AI algorithms provides novel solutions for massive MIMO systems involving many antennas and beam arrays. A beam codeword consists of analogue phase-shifted values and is applied to the antenna elements to form an analogue beam in \cite{9034044}, base beam selection with deep learning (DL) algorithms are proposed for using channel state information for the sub-6 GHz links. In addition to beam prediction, location and size of vehicles information are used to predict the optimal beam pair \cite{8644288}. Location-based beamforming solutions are more suitable for line-of-sight (LOS) communication. On the other hand, the same locations with the non-line-of-sight (NLOS) transmission need different beamforming solutions. 

In the literature, most studies have focused on the communication methods to increase cellular technologies' performance but usually ignore the security and privacy issues and the integration of currently emerging AI tools into 6G. It is expected that 6G networks would provide better performance than 5G ones and satisfy emerging services and applications. Authors in \cite{8766143} and introduce AI as a critical enabler and offer a comprehensive review of 6G networks, including usage scenarios, requirements, and promising technologies for 6G networks. The review paper indicated that the promising technologies such as blockchain-based spectrum sharing and quantum communications and computing could significantly improve the 6G’s spectrum in terms of efficiency and security while comparing conventional techniques. The study \cite{8808168} discusses the key trends and AI-powered methodologies for 6G network design and optimization. Authors in \cite{9240722} introduce and analyze the key technologies and application scenarios brought by the 6G networks. However, there’s also reason to be concerned about security risks. 6G carries over and introduces new risks, which must be addressed to ensure its secure and safe use. The study \cite{lu2020security} addresses the fundamental principles of 6G security, discusses significant technologies related to 6G security, and presents several issues regarding 6G security. Authors in \cite{wang2020security} investigate the fundamental security and privacy challenges associated with each key technology and potential applications, i.e., real-time intelligent edge, distributed AI, intelligent radio, and 3D intercoms, for 6G networks. The study in \cite{chorti2021context} proposes a framework incorporating context-awareness in quality of security (QoSec) that leverages physical layer security (PLS) for 6G networks. The framework identifies the security level required and proposes adaptive, dynamic, and risk-aware security solutions. The key component of 6G is the integration of AI, i.e., self-learning architecture based on self-supervised algorithms, to be able to improve the performance of the network for tomorrow's wireless cellular systems \cite{9311932}. It is expected that a secure AI-powered structure can protect privacy in 6G. However, AI itself may be attacked or abused, resulting in privacy violations. The authors in \cite{kuzlu2021role} also indicate that some attackers simply can replace a legitimate model with an already poisoned model prepared ahead of time, i.e., attacking beneficial AI in such a way that the AI works against itself. The study in \cite{9146540} provides a comprehensive survey of DL and privacy in 6G, with a view to further promoting the development of 6G and privacy protection technologies. With the use of DL algorithms in 6G’s physical layer functions, such as channel estimation, modulation recognition, and channel state information (CSI) feedback, the physical layer faces new challenges caused by an adversarial attack. The authors in \cite{9143575} investigate the impact of possible adversarial attacks on DL-based CSI feedback. According to the results, an adversarial attack may cause a destructive effect on DL-based CSI feedback, and transmitted data can be easily tampered with adversarial perturbation by malicious attackers due to the broadcast nature of wireless communication.

Smart cities has ushered in an era in which everyday objects, such as cars, toasters, window blinds and even toothbrushes, can be connected to the Internet. Indoor-IoT (i-IoT) sector makes up more than half of the total IoT market. Indoor is where most of the time people live and work. Moreover, the majority of the unresolved grand technical challenges hindering the wider adoption of IoT are found in indoor deployments. These include: 1) Non-standardised indoor environment, e.g., homes, devices come from different vendors but still need to interoperate to achieve common goals. 2) i-IoT solutions are often managed by IT novice users (owners of the system), thus systems must be able to self-recover from failures quickly and cost-effectively. 3) IoT devices are commonly installed with a number of security vulnerabilities which renders them as a backdoor to hack home/corporate networks. 4) i-IoT devices are often connected to actuators meaning that a successful attack could result in physical harm or risk, e.g., privacy risk through viewing CCTV camera. i-IoT will connect diverse range of devices regardless of their vendor, communication technology or software/hardware platform. The current i-IoT space is largely dominated by vertical industries, each with domain-specific solution that does not interoperate with others. This lack of interoperability has limited the development of i-IoT applications. In addition to security and privacy, another key challenge in i-IoT is how to adaptively manage the avalanche of heterogeneous devices and intelligently deliver smart city services on a consistent QoS basis.  Different architectural models, such as the three- and five-layer models \cite{6002149,6424332,7123563}, have been studied in literature. The outcome can be likened to a home in which heating, ventilation and air conditioning, TV, audio system, security system, lighting, etc., each has its remote control, but no single one can control all devices; in other words, there is currently a lack of unified reference architecture that addresses the specific needs of i-IoT networks. Some previous EU projects such as IOT-A-257521, iCore-287708, BUTLER-287901, COMPOSE-317861 and FIESTA-IoT-643943 have investigated or trialled different aspects of IoT including: interoperability, location/context awareness, cognitive networking and service composition. Other recent projects, such as SENSEI-610916, ASPIRE-215417 and SPITFIRE-258885, targeted the service layer. The perspectives developed in these projects were based mainly on the specific requirements of the corresponding use cases and are not crosscutting. For example, none of the previous studies or projects considered the heterogeneous communications technologies and how to integrate them for i-IoT applications. Furthermore, i-IoT networks need to be self-aware and adaptive, i.e., they should be able to learn users’ behaviour and act intelligently and proactively on behalf of the users; this self-awareness and adaptiveness still remains largely unexplored to date.
There is little discussion in literature on the next step in the evolution of i-IoT, which is to create a living, intelligent, flexible and dynamic i-IoT that supports autonomous network reconfiguration to provide distributed management through analysing control data in real-time to deliver the insights user/business need. Past research and development efforts, e.g., \cite{RAKOTONIRAINY201678,rahman2018enabling,vermesan2017internet}, explored the applications of ML-based techniques to improve the communication infrastructure performance and support existing services. However, these representative examples of previous work show that such efforts have addressed various segments of the overall network control/management optimisation problem. is-Smart approach is to unlock the potential of ML techniques, data analytics and Software Defined Networking (SDN) to foster the development of a coherent intelligent network control solution, which addresses the increasingly dynamic and responsive i-IoT. This solution integrates reasoning and problem-solving techniques, computational analysis and predictive algorithms to operate i-IoT effectively, as it continues to grow and evolve, to achieve optimal user experience (QoS).
Smart cities enabled i-IoT systems presents large attack surfaces. Recently, there have been many instances of cyber-attacks, which have resulted in the breach of sensitive data and even physical damage. For instance, in 2016, Oracle Dyn company was targeted by the Mirai attack \footnote{https://www.oracle.com/cloud/networking/dns/}. The following years, Mirai variants’ affected several companies \cite{7971869}.
There are many studies on IoT security based on the traditional security techniques such as secure communication, authentication, authorisation and access control \cite{AMMAR20188}. However, traditional security techniques often involve heavy computation and processing and, therefore, require a large amount of memory and battery power to achieve high security. These requirements contradict with the nature of the IoT constraints \cite{8378034}. Improved techniques of providing security are, therefore, required to guarantee the security of IoT systems \cite{8319238} in particular for countering new threats such as large-scale botnet attacks.
One of the essential obstacles facing IoT researchers belongs to preparing and processing huge amounts of data \cite{MAHDAVINEJAD2018161}. Accordingly, ML and Data Mining (DM) are widely used to improve the performance of cyber-attack detection and prevention systems \cite{shanbhogue2017survey}, and to increase the security of transmitting sensitive data to public cloud. An integration between Multilayer Perceptron Neural Networks (MLP) and Particle Swarm Optimization Algorithm (PSO) was employed in \cite{saljoughi2017attacks}. 

To sum up, the integrating the DL algorithms for the 6G and beyond technologies leads to potential security problems. Mainly, most of the studies are focus on building DL algorithms for the 6G communication problems and ignore the security concerns. In \cite{8395149} the authors presented promising results for the mmWave beam prediction for several base stations (BSs) with multiple users by using DL algorithms for different environmental scenarios. On the other hand, the most of the proposed DL methods cannot work under attack. The beamforming method may seem naturally secure because it transmits signals to desired directions. More specifically, security can be further guaranteed from the viewpoint of physical security when the beamforming vector is appropriately designed. On the other hand, in DL models, the malicious user can penetrate the legitimate users' device or generate a copy of the users' signal to impersonate users with malicious software. For this, the security concerns for DL models for wireless communication are different from the traditional wireless communication systems. Based on the shortcomings of the literature with regards to security issues, in this paper, we deal with the security problem of DL application for beamforming prediction. 

We consider two research questions: i) Is the proposed DL-based mmWave beam prediction model vulnerable for the adversarial attacks, ii) Is iterative adversarial training able to mitigate adversarial attacks. First, we implemented a beam prediction algorithm using a DL model to answer these questions. Secondly, we attack the beam prediction algorithm with the Fast-Gradient Sign Method (FGSM), an essential and powerful attack for DL models. Eventually, we compare the mean square error (MSE) values of the undefended DL model and attack the DL model with FGSM. The MSE value increases about 40.14 times higher with the attack. Thirdly, we proposed an adversarial training-based mmWave beam prediction model to protect the model against FGSM adversarial machine learning (ML) attacks. In addition to the beam prediction, our new DL model learns the attack noise injection patterns and trains itself with manipulated input data, denoted as adversarial training.

\subsection{Contributions}
This study aims to make a more secure DL-based mmWave beam prediction against attack for the DL model. The future of wireless communication  must consider using AI models. The attack against AI models is different from well-known wireless physical layer security achieved by exploiting the properties of the physical layer, as the name suggests, such as interference, thermal noise, channel information, jamming, etc. The purpose of the attack on wireless physical layer security is to make the transmitted signal non-predictive to decrease the secrecy capacity. In this way, the legitimate users could not demodulate the transmitted signal. On the other hand, the purpose of the attacks against the DL models to manipulate the transmitted data. The attacker imitates the legitimate user. In this case, we develop an attack model for the base station to mimic the user's transmitted signal.

The principles of traditional wireless communication system models and DL-based models are different. Wireless communication systems are not prediction or approximation-based. On the other hand, a typical DL model trains its neuron weights to extract the relationship between the input and output of a system. The resulting predictive model is defined as the decision boundaries of the input data. The final decision boundaries are nonlinear lines that separate the outputs, unlike the traditional wireless communication model. For instance, at the energy detection model \cite{1447503}, the input values compare with a threshold value to decide the false and true regions. Here, the boundary of the false and true region is a linear function. Therefore, it is possible to detect a slight change in the input signal. On the other that, if the predicted boundary is nonlinear, it causes vulnerabilities. It is open for adversarial ML attacks. We choose the most common and powerful attacking method for the DL model that is FGSM. This attack model maximizes the lost values of the classifier by adding a modest noise vector. While the traditional FGSM attack only uses real numbers to manipulate data, we modified the FGSM attack model to change the transmitted signal's amplitude and phase values with complex numbers. Thus, our main contributions for this paper are listed as below:

\begin{itemize}
    \item We show that an undefended DL-based mmWave beam prediction model system is vulnerable against carefully designed adversarial noise.
    \item We modified the FGSM attack to manipulate the transmitted signal in the complex domain for amplitude and phase values. After the attack, the system achievable rate performance became inoperable.
    \item We trained an undefended DL-based mmWave beam prediction model by adversarial training with the FGSM attack. Therefore, the system achievable rate performance became very close to the undefended model without attack. 
\end{itemize}

We implemented the proposed model with three scenarios; outdoor, outdoor with LOS and blocked users, and indoor environment. Each scenario is executed under three cases that undefended, undefended under attack and secure model.
\subsection{Organization}
The rest of the paper is organized as follows: Section \ref{sec:preliminaries} describes background information about adversarial machine learning and adversarial training-based mitigation methods. The Section \ref{sec:system_overview} shows our system overview. Section \ref{sec:experiments} evaluates the proposed mitigation method for DL based mmWave beam prediction vulnerabilities, and the section \ref{sec:conclusion} concludes this paper.

\section{Preliminaries} \label{sec:preliminaries}
\subsection{Using Machine Learning Models to estimate RF beamforming vectors}
Using the benefits of DL algorithms gives a novel solution for a massive MIMO channel training and scanning a large number of narrow beams. The beams depend on the environmental conditions, like user and BSs locations, furniture, trees, buildings etc. It is very difficult to define all these environmental conditions as a closed-form equation. A good alternative is to use omni and quasi-omni beam patterns to predict the best RF beamforming vectors. We are using these beam patterns benefits to consider the reflection and diffraction of the pilot signal. This research uses the DL models for mmWave beam prediction in \cite{8395149}, thanks to their mathematical calculations.

The DL solution consists of two states: training and prediction. Firstly, the DL model learns the beams according to the omni-received pilots. Secondly, the model uses the trained data to predict the RF beamforming vector for the current condition. 

\subsubsection{Training Steps}
The user sends uplink training pilot sequences for each beam coherence time $T_{B}$. BSs combine received pilot sequences on RF beamforming vector and feed them to the cloud. The cloud uses the received sequences from all the BSs as the input of the DL algorithm to find the achievable rate in (\ref{Eqn:b6}) for every RF beamforming vector to represent the desired outputs,
\begin{equation}\label{Eqn:b6}
	{R}_{n} ^{(p)}= \mathrm{arg max} \frac{1}{K} \sum_{n=1}^{N}  \log_2  \left(  1+ SNR\vert \textbf{h}_{k,n}^T \textbf{g}_{p} \vert^2  \right)
\end{equation}
where $\textbf{g}_{p}$ is the channel coefficient for omni-beams, and $\textbf{h}_{k,n}$ is channel coefficient for $n$-th BS at the $k$-th subcarrier.
\subsubsection{Learning Steps} 
In this stage, the trained DL model is used to predict the RF beamforming vectors. Firstly, the user sends an uplink pilot sequence. The BSs combine these sequences and send them to the cloud. Then, the cloud uses the trained DL model to predict the best RF beamforming vectors to maximize the achievable rate for each BS. Finally, BSs use the predicted RF beamforming vectors to estimate the effective channel $\textbf{h}_{k,n}$.
\subsection{Attack to Machine Learning Algorithms: Adversarial Machine Learning} 
To sum up, adversarial machine learning is an attack technique that attempts to fool neural network models by supplying craftily manipulated input with a slight difference \cite{2016arXiv161101236K}.

\begin{figure}[h]
    \centering
    \includegraphics[width=0.5\linewidth]{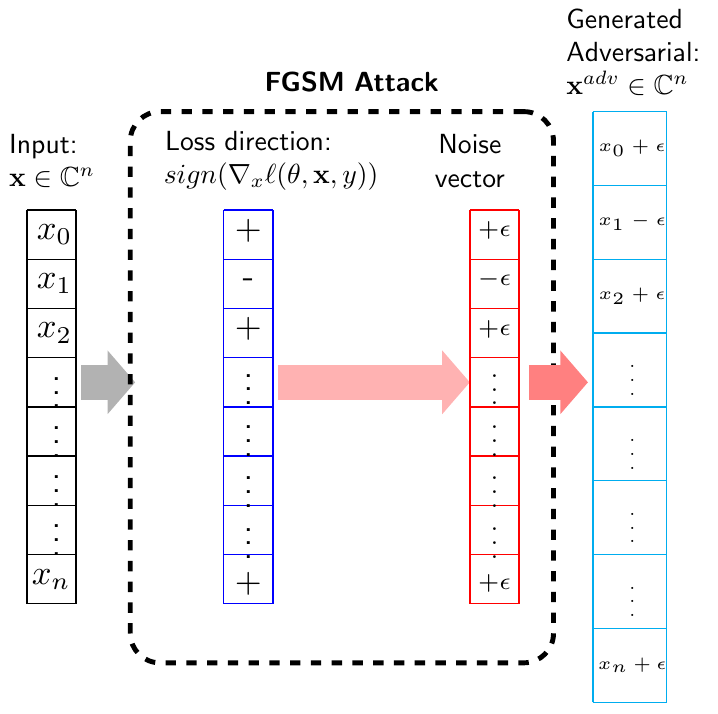}
    \caption{FGSM attack steps. The input vector $\mathbf{x} \in \mathbb{C}^n$ is poisoned with loss maximization direction. }
    \label{fig:fgsm_detail}
\end{figure}

Attackers apply model evasion attacks for phishing attacks, spams, and executing malware code in an analysis environment \cite{8965459}. There are also some advantages to attackers in misclassification and misdirection of models. In such attacks, the attacker does not change training instances. Instead, he tries to make some small perturbations in input instances in the model's inference time to make this new input instance seem safe (i.e., normal behaviour) \cite{2021arXiv210204150F}. We mainly concentrate on this kind of adversarial attack in this study. There are many attacking methods for DL models, and the FGSM is the most straightforward and powerful attack type. We only focus on the FGSM attack, but our solution to prevent this attack can be applied to other adversarial machine learning attacks. FGSM works by utilizing the gradients of the neural network to create an adversarial example to evade the model. For an input instance $\mathbf{x}$, the FGSM utilizes the gradients $\nabla_x$ of the loss value $\ell$ for the input instance to build a new instance $\mathbf{x}^{adv}$ that maximizes the loss value of the classifier hypothesis $h$. This new instance is named the adversarial instance. We can summarize the FGSM using the following explanation:
\begin{equation}
	\mathbf{x}^{adv} = \mathbf{x} + \epsilon \cdot sign(\nabla_x \ell(\mathbf{\theta},\mathbf{x},y))
\end{equation}
By adding a slowly modest noise vector $\eta \in \mathbb{R}^n$ whose elements are equal to the sign of the features of the gradient of the cost function $\ell$ for the input $\mathbf{x} \in \mathbb{R}^n$, the attacker can easily manipulate the output of a DL model.
The Figure \ref{fig:fgsm_detail} shows the details of the FGSM attack.

Attackers can get involved in the system by using different way such as mobile malware applications, copying mobile base stations. An attacker can can be a device or an application in addition to a human. Attack transferability achieves an attack against a DL model to be valid against a different, unknown model. In the attack transferability paradigm, one attacker can build another DL model to extract the decision boundaries of the input data. Then the attacker can create the adversarial samples using his own model's vulnerabilities. Experimental confirmation for attack transferability has been shown in recent works \cite{2018arXiv180902861D}.

\subsection{Attack to Training Steps: Adversarial Training}
Adversarial training is a widely recommended defense technique that implies generating adversarial instances using the gradient of the victim classifier and then re-training the model with the adversarial instances and their respective labels. This technique has been demonstrated to be efficient in defending models from adversarial attacks. 

Let us first think of a common classification problem with training instances $X \in \mathbb{R}^{m \times n}$ of dimension $d$, and a label space $Y$. We assume the classifier $h_\theta$ has been trained to minimize a loss function $\ell$ as follows:
\begin{equation}
	\label{eq:cost-func}
	\underset{\theta}{min}\frac{1}{m} \sum_{i=1}^{m} \ell(h_\theta(\mathbf{x}_i,y_i))
\end{equation}
Given a classifier model $h_\theta(\cdot)$ and an input instance $x$ with a responding output $y$, then an adversarial instance $x^{adv}$ is an input such that:
\begin{equation}
	\label{eq:adv_ex}
	h_\theta(x^{adv}) \neq y \,\,\,\,\, \wedge \, d(x,x^{adv}) < \epsilon
\end{equation}
where $d(\cdot,\cdot)$ is the distance metric between two input instances, the original input $x$ and the adversarial version $x^{adv}$. Most actual adversarial model attacks transform Eq. (\ref{eq:adv_ex}) into the following optimization problem:
\begin{equation}
	\underset{x}{arg\,max} \, \ell \left(h_\theta(x^{adv}),y\right)
\end{equation}
\begin{equation}
	s.t. \,\,\,\,\, d(x,x^{adv}) < \epsilon
\end{equation}
where $\ell$ is the loss function between predicted output $h(\cdot)$ and correct label $y$.
In order to mitigate such attacks, at per training step, the conventional training procedure from Eq. (\ref{eq:cost-func}) is replaced with a \texttt{min-max} objective function to minimize the expected value of the maximum loss, as follows:
\begin{equation}
	\underset{\theta}{min} \, \underset{(x,y)}{\mathbb{E}} \left(\underset{d(x,x^{adv})<\epsilon}{max} \ell(h(x^{adv}),y) \right)
\end{equation}

\section{System Model}\label{sec:system_overview}
\begin{figure}[t!]
    \centering
    \includegraphics[width=0.5\linewidth]{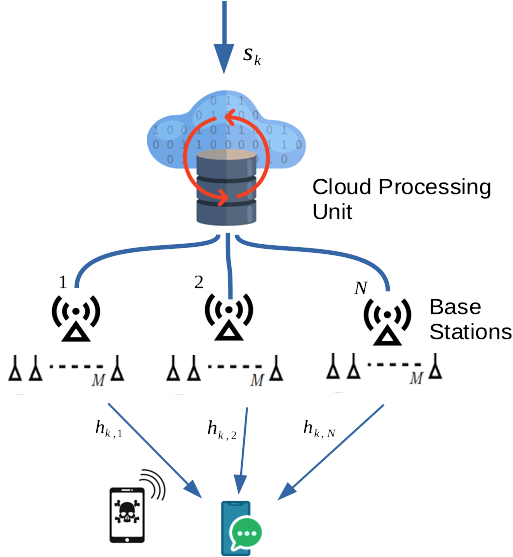}
    \caption{Block diagram of the mmWave beamforming system.}
    \label{fig:downlink_comm}
\end{figure}

mmWave communication system employs a massive amount of antennas with beamforming to control a wave-front direction by weighting the magnitude and phase in each antenna. We assume that each BS has one RF chain to provide analogue beamforming architecture that is not as expensive and complex as the other approaches in \cite{6979963}. The mmWave communication system model is given in Figure  \ref{fig:downlink_comm}. Here, $N$ is the number of BSs serving one mobile user with an equipped single antenna. A centralized/cloud processing unit is used to connect all BSs and processing.

The downlink received signal at $k$-th subcarrier is expressed as 
\begin{equation}\label{Eqn:b2}
	\textbf{y}_{k} =\sum_{n=1}^{N} \textbf{h}_{k,n}^T \textbf{x}_{k,n} + v_k 
\end{equation}
where $\textbf{h}_{k,n}$ denotes the channel vector between $n$-th BS and the user. $v_k$ is additive white Gaussian noise (AWGN) with variance $ \sigma ^{2} $, i.e., $N(0, \sigma^2)$ for $k$-th subcarrier. Here, $\textbf{x}_{k,n}$ is transmitted complex baseband signal from the $k$-th subcarrier and $n$-th BS is given as
\begin{equation}\label{Eqn:b1}
	\textbf{x}_{k,n} = \textbf{f}_{n}c_{k,n}s_k 
\end{equation}
where $s_k$ is the data symbol $  \textbf{s} = [s_1, s_2,\dots, s_K]$ with $K$ subcarriers is firstly precoded by code vector $\textbf{c}_{k,n}= [c_{k,1}, c_{k,2},\dots, c_{k,N}]^T $ at each subcarrier on each base station. Then, every BS applies analog beamforming with beam steering vector $\textbf{f}_{n}$  to obtain downlink transmitted signal $\textbf{x}_{k,n}$. Beam steering vector defines for each BS antenna as [$\textbf{f}_{n}]_m =  \frac{1}{\sqrt{M}}e^{j\theta_{n,m}}$ where $\theta_{n,m}$ is a quantized angle. 

To support mobile users, beamforming vectors are recalculated constantly beamforming vectors within channel coherence time, denoted $T_{C}$ which depends on user mobility and channel multi-path components. Also, the beams stay aligned on beam coherence duration, denoted $T_{B}$, and $T_{C}$ is usually shorter than $T_{B}$ \cite{7742901}. The time duration of $T_{B}$ decreases for the users with higher mobility that causes to lower data rate for the same beamforming vectors and beam training overhead. Thus, the effective achievable rate is defined as follow.
\begin{equation}\label{Eqn:b3}
R_{eff} =
   \left( 1- \frac{T_{TR}}{T_{B}} \right) 
   \sum_{k=1}^{K}  \log_2  \left(  1+ SNR \left \vert  \sum_{n=1}^{N}   \textbf{h}_{k,n}^T \textbf{f}_{n}c_{k,n}  \right \vert^2  \right). 
\end{equation}

Here, the beamforming vectors are redesigned in each first training time $T_{TR}$ in beam coherence time, $T_{B}$ and the rest of it is used for the data transmission by using the redesigned beamforming vectors.

\subsection{Modified FGSM}

The traditional FGSM attack succeeds in classification models whose input type is real numbers ($\mathbf{x} \in \mathbb{R}^m $). However, on the other hand, the data used in the field of communication systems consist of complex numbers ($\mathbf{x} \in \mathbb{C}^m $) with a real and an imaginary part. For this reason, the FGSM attack needs to be modified for inputting complex numbers in mmWave estimation. For this purpose, we updated the FGSM attack as shown in Algorithm \ref{alg:complex-fgsm} to be used in 6G beyond technologies. 

\begin{algorithm}
\SetAlgoLined
\DontPrintSemicolon
\KwIn{$\mathbf{x}  \in \mathbb{C}^m, \mathbf{y} \in \mathbb{R}^n, F, \epsilon, \alpha$}
\KwOut{$\mathbf{x}_{t+1}$}
\tcc{convert $\epsilon \in \mathbb{R}$ from real number domain to $\epsilon\_complex \in \mathbb{C}$ complex domain where $Re\{\epsilon\_complex\} = \epsilon$ and $Im\{\epsilon\_complex\} = \epsilon$}
$\epsilon\_complex \gets (\epsilon + \epsilon \cdot j)$\;
$\mathbf{x}_0 \gets \mathbf{x}$\;
\While{$n < N$}
{
\tcc{update $\mathbf{x}$ using the loss direction}
$\mathbf{x}_{(t+1)} = clip_{\mathbf{x}, \epsilon}(\mathbf{x}_t + \epsilon\_complex \cdot sign(\nabla_\mathbf{x} \ell(\mathbf{x}_t,F,\mathbf{y}))) $ \;
\tcp{If the distance between manipulated input's prediction and real output is greater than $\alpha$}
\If{$distance_{Euclidian} (F(\mathbf{x}_{t+1}) -\mathbf{y}) \geq \alpha$}
{
    end while\;
}
}
\Return $\mathbf{x}_{t+1}$\;
\caption{Algorithm for FGSM (complex numbers based)  $\mathbf{x} \in \mathbb{C}^n$ is the benign input, $F$ is the model function learnt during training, $N$ is the number of iterations, $\alpha$ is the maximum allowed perturbation, $\epsilon$ is the step size, }
\label{alg:complex-fgsm}
\end{algorithm}

\subsection{Adversarial Training}
\begin{figure}[htbp!]
    \centering
    \includegraphics[width=0.9\linewidth]{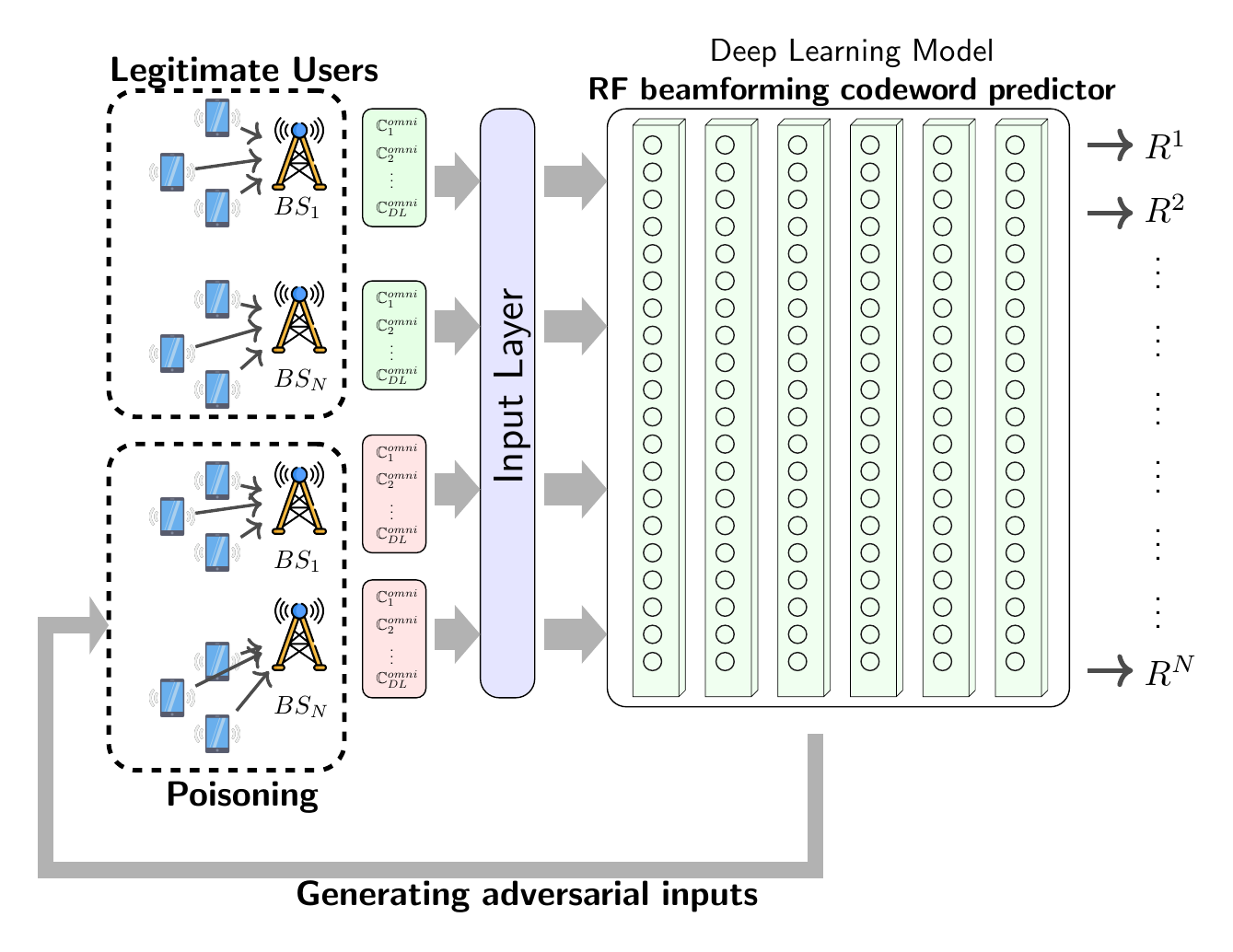}
    \vspace{-5pt}
    \caption{The diagram of RF beamforming codeword adversarial training.}
    \label{fig:deepmimo_adv_learning}
\end{figure}

Figure \ref{fig:deepmimo_adv_learning} shows the adversarial training process. After the model is trained, adversarial inputs are created using the model itself, combined with legitimate users' information and added to the training. When the model reaches the steady-state state, the training process is completed. In this way, the model will both predict RF beamforming codeword for legitimate users while at the same time being immune to the craftily designed noise attack that will be added as input.
\subsection{Capability of the Attacker}
We assumed that the attacker's primary purpose is to manipulate the RF model by applying carefully crafted noise to the input data.  In a real-world scenario, this white-box setting is the most desired choice for an attacker that does not take the risks of being caught in a trap. The problem is that it requires the attacker to access the model from outside to generate adversarial examples. After manipulating the input data, the attacker can exploit the RF beamforming codeword prediction model's vulnerabilities in the same manner as in an adversary's sandbox environment. The prediction model predicts the adversarial instances when the attacker can convert some model's outputs to other outputs (i.e., wrong prediction).

\begin{figure}[htbp!]
\centering
\includegraphics[width=0.8\linewidth]{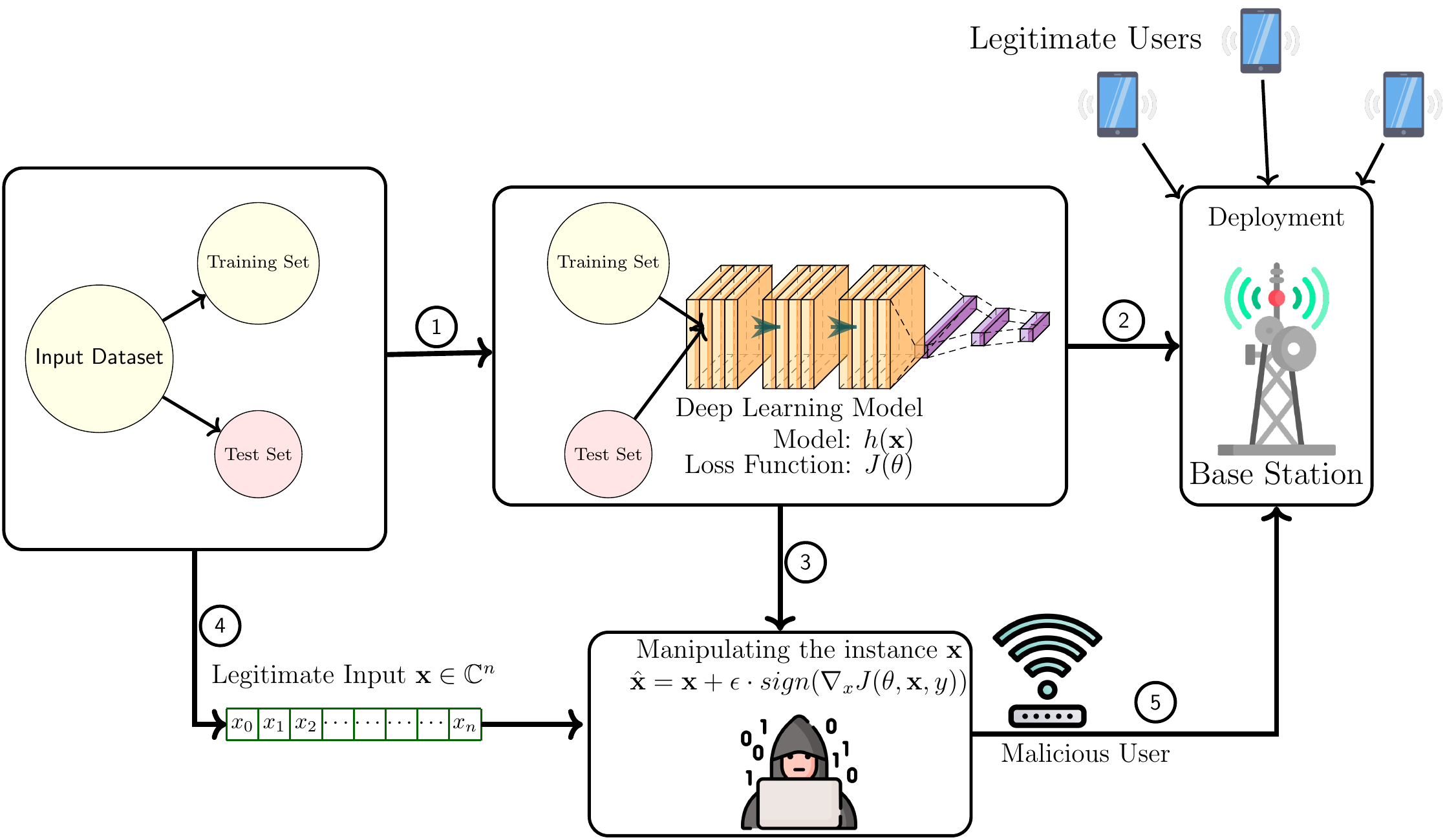}
\caption{RF Beamforming manipulation process.}
\label{fig:adv-ml}
\end{figure}

However, to prevent this noise from being easily noticed, the attacker must answer an optimization problem to determine which regions in the input data (i.e., beamforming) must be modified. By solving this optimization problem using one of the available attack methods \cite{8965459}, the attacker aims to reduce the prediction performance on the manipulated data as much as possible. In this study, to limit the maximum allowed perturbation allowed for the attacker, we used $l_\infty$ norm, which is the maximum difference limit between original and adversarial instances. Figure \ref{fig:adv-ml} shows the attack scenario. The attacker gets an legitimate input, $\mathbf{x}$, creates a noise vector with an $\epsilon$ budget $\eta = \epsilon \cdot sign(\nabla_x \ell(\mathbf{\theta},\mathbf{x},y))$, sums the input instance and the craftily designed noise to create adversarial input $\mathbf{x}^{adv} = \mathbf{x} + \eta$.

\section{Experiments}\label{sec:experiments}
\begin{figure}[htbp!]
    \centering
    \includegraphics[width=0.7\linewidth]{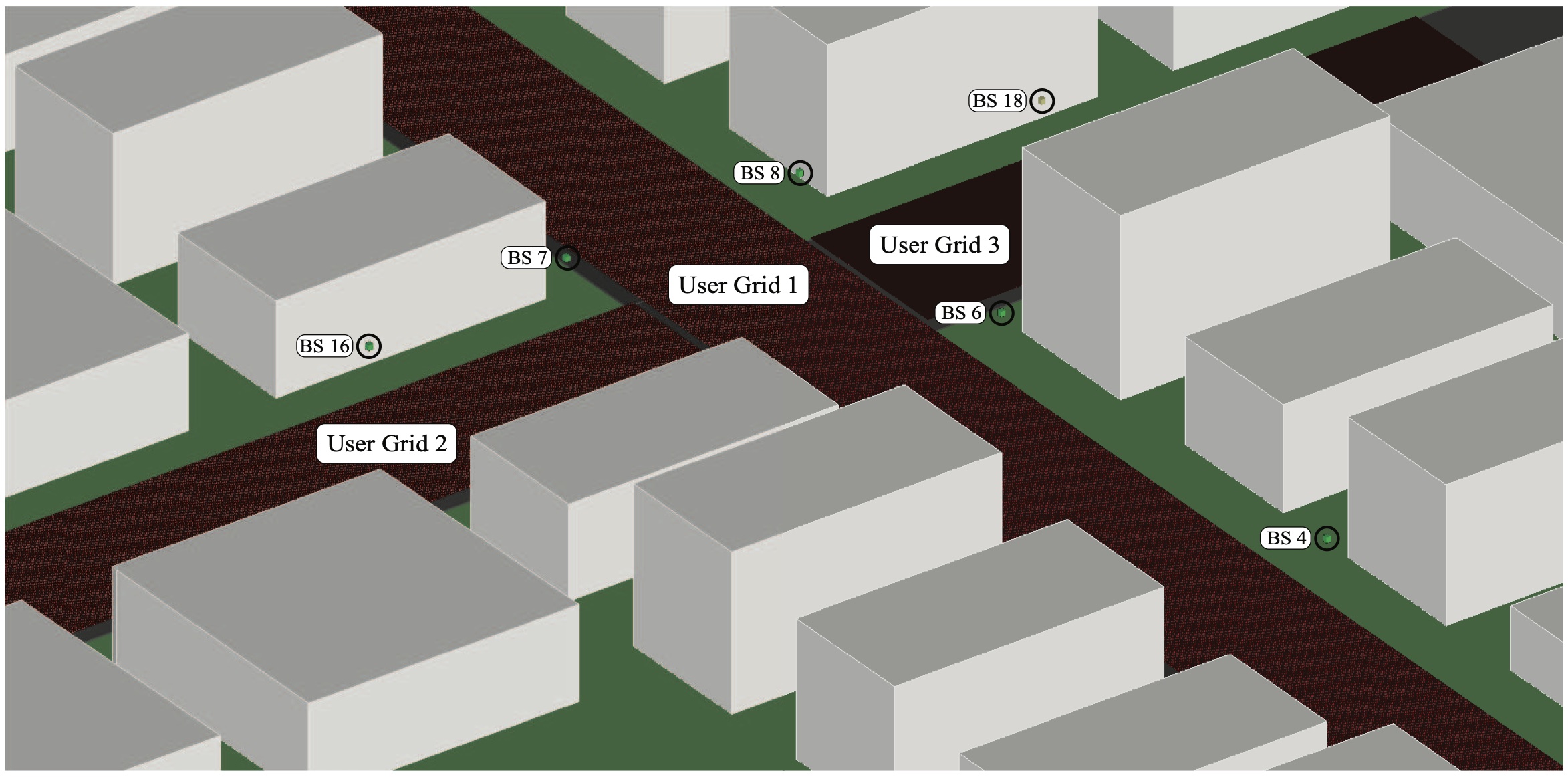}
    \caption{Scenario 1- Outdoor scenario \cite{deepmimo_adam}. }
    \label{fig:scenario}
\end{figure}

In the experiments, we tested our model with three different cases for three different scenarios. The cases are given as: 
\begin{itemize} 
    \item \textbf{Case 1:} Undefended model: We implement undefended DL-based mmWave beam prediction model which is vulnerable to attacks. 
    \item \textbf{Case 2:} Undefended model under FGSM attack: We attack with FGSM to undefended model to obtain an achievable rate of the  DL model under attack. It is the worst case of the model that needs to be overcome.
    \item \textbf{Case 3:} Secure model: DL-based mmWave beam prediction model is adversarial trained  against FGSM attack.
\end{itemize}
The outcomes of these three cases allow us to compare the model performance under attack with undefended case and secure case. Also, we implemented the proposed model for different scenarios including outdoor and indoor scenarios with the details below \cite{deepmimo_adam}:

\textit{Scenario 1- Outdoor scenario:} This is an outdoor scenario of two streets with an intersection as shown in Figure \ref{fig:scenario}. The scenario includes 18 BSs with  16 $\times$ 16 uniform planar array (UPA) and uniformly distributed more than one million users with a single dipole antenna in 3 user grids. The operating frequency is 60 GHz.

\begin{figure}[htbp!]
    \centering
    \includegraphics[width=1.0\linewidth]{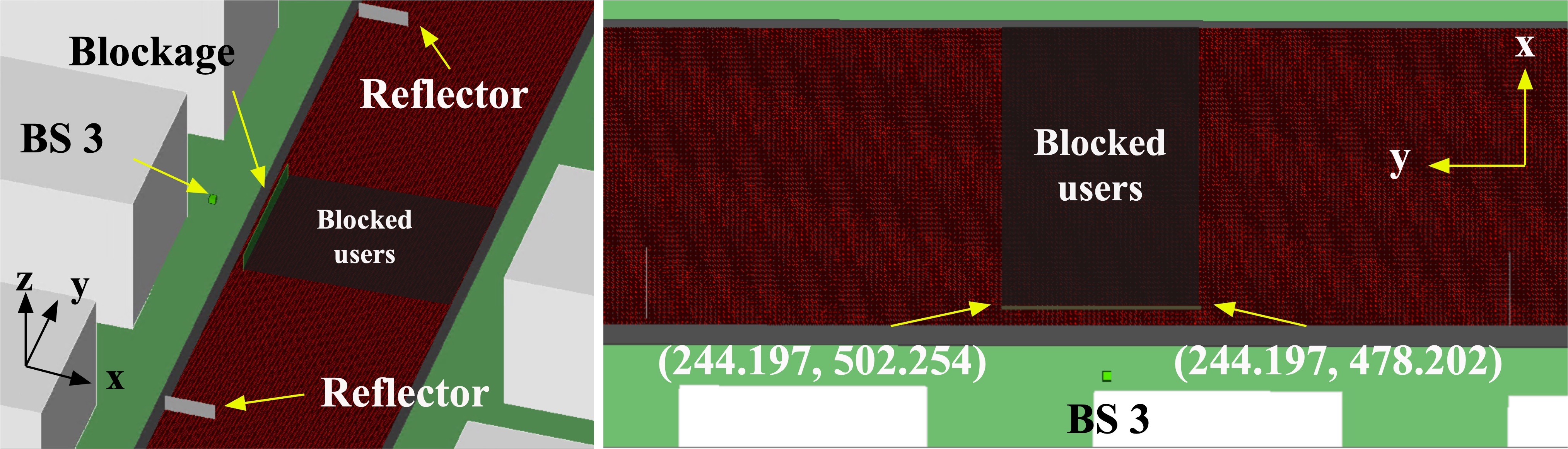}
    \caption{Scenario 2- Outdoor scenario with LOS and blocked users \cite{deepmimo_adam}.}
    \label{fig:O1_28B_V2}
\end{figure}

\textit{Scenario 2- Outdoor scenario with LOS and blocked users:} It is also an outdoor scenario with LOS and blocked users is given in Figure \ref{fig:O1_28B_V2}. There is a single BS with LOS connections with some users and NLOS connections with other users.
The operating frequency is 3.5 GHz.

\begin{figure}[htbp!]
    \centering
    \includegraphics[width=0.5\linewidth]{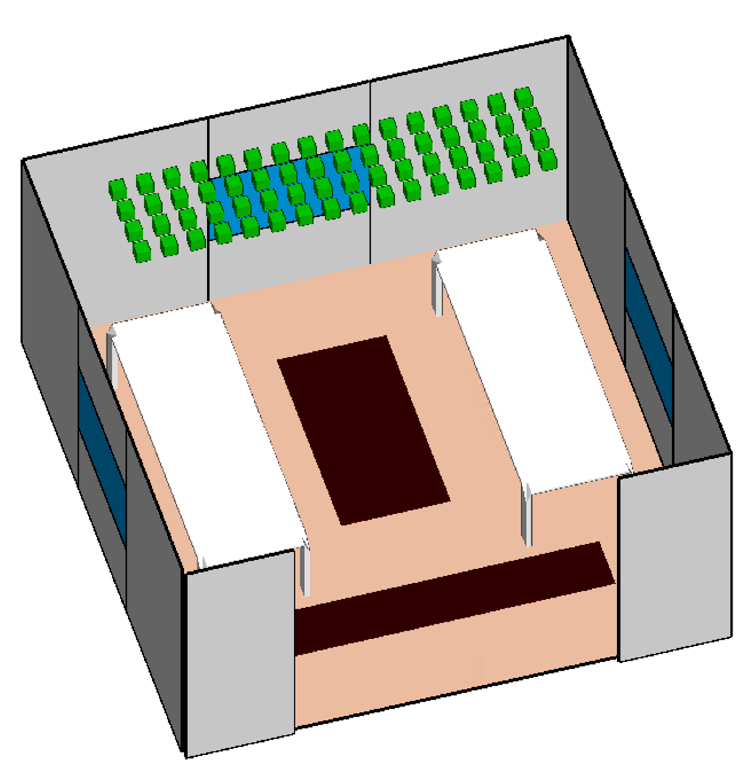}
    \caption{Scenario 3- Indoor scenario with distributed massive MIMO \cite{deepmimo_adam}.}
    \label{fig:O1_28B_V3}
\end{figure}

\textit{Scenario 3- Indoor scenario with distributed massive MIMO:} It is for indoor $10 m \times 10 m$ room scenario and 64 antennas tiling up part of the ceiling at the height of 2.5 m from the floor that is given in Figure \ref{fig:O1_28B_V3}. The operating frequencies are 2.4 GHz and 2.5 GHz.

Our motivations in form these cases is to maximize the system effective achievable rate for the system under attack in Eq. (\ref{Eqn:b3}). We performed the experiments using the Python scripts and ML libraries: Keras, Tensorflow, and Scikit-learn, on the following machine: 2.8 GHz Quad-Core Intel Core i7 with 16GB of RAM. For all scenarios, two models, undefended and adversarial trained, were built to obtain prediction results. In the first model, the model is trained without any input poisoning. The first model (i.e., undefended model) was used with legitimate users (for C1) and adversaries (for C2). The second model (i.e., the adversarially trained model) was used under the FGSM attack. The hyper-parameters such as the number of hidden layers and the number of neurons in the hidden layers, the activation function, the loss function, and the optimization method are the same for both models.

The model architectures and selected hyper-parameters are given in Table \ref{tab:dl_arch} and in Table\ref{tab:dl_arch_params} respectively.
\begin{table}[ht!]
\centering
\caption{Model architecture}
\label{tab:dl_arch}
\begin{tabular}{c|c}
	\hline \textbf{Layer type} & \textbf{Layer information} \\ \hline
	Fully Connected + ReLU & 100 \\
	Fully Connected + ReLU & 100 \\
	Fully Connected + ReLU & 100 \\
	Fully Connected + TanH & 1 \\ \hline
\end{tabular}
\end{table}

\begin{table}[htbp!]
\centering
\caption{Milimater-wave beam prediction model parameters}
\label{tab:dl_arch_params}
\begin{tabular}{c|c} 
	\hline
	\textbf{Parameter} & \textbf{Value} \\
	\hline 
	Optimizer & Adam \\
	Learning rate & 0.01 \\
	Batch Size & 100 \\
	Dropout Ratio & 0.25 \\
	Epochs & 10 \\
	\hline
\end{tabular}
\end{table}

\subsection{Research Questions}\label{RQ}
We consider the following two research questions (RQs):
\begin{itemize}
    \item \textbf{RQ1}: Is the DL-based RF beamforming codeword predictor vulnerable for adversarial machine learning attacks?
    \item \textbf{RQ2}: Is the iterative adversarial training approach a mitigation method for the adversarial attacks in beamforming prediction?
\end{itemize}
 \subsection{RF Beamforming Data Generator}\label{sec:datagenerator}
We employed the generic DL dataset for millimeter-wave and massive MIMO applications (DeepMIMO) data generator in our experiments \cite{alkhateeb2019deepmimo}. 

This section conducts experiments on the mmWave communication and massive MIMO applications dataset from the publicly available data set repository. We implemented the proposed mitigation method using Keras and TensorFlow libraries in the Python environment. 

\subsection{Results for RQ1}\label{sec:res_rq1}
\begin{figure}[h]
    \centering
    \includegraphics[width=0.7\linewidth]{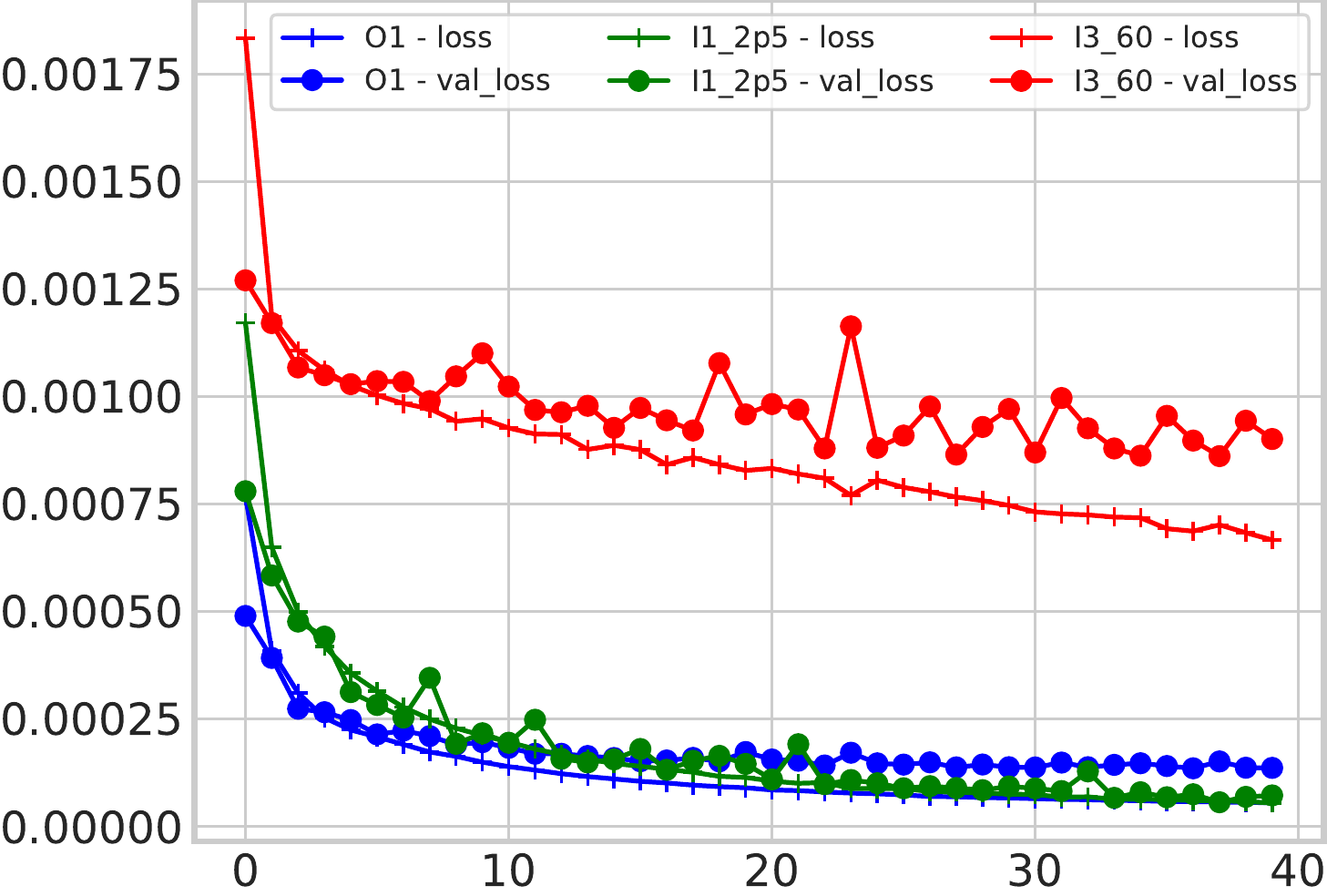}
    \caption{The beamforming prediction model training history.}
    \label{fig:history_org}
\end{figure}

Figure \ref{fig:history_org} shows the training history of the beamforming prediction model with 35.000 training instances. The model is trained with clean (i.e. non-perturbated) instances.

\begin{figure}[!htbp]
    \centering
    \begin{subfigure}[b]{0.49\linewidth}
         \centering
         \captionsetup{justification=centering}
         \includegraphics[width=1\linewidth]{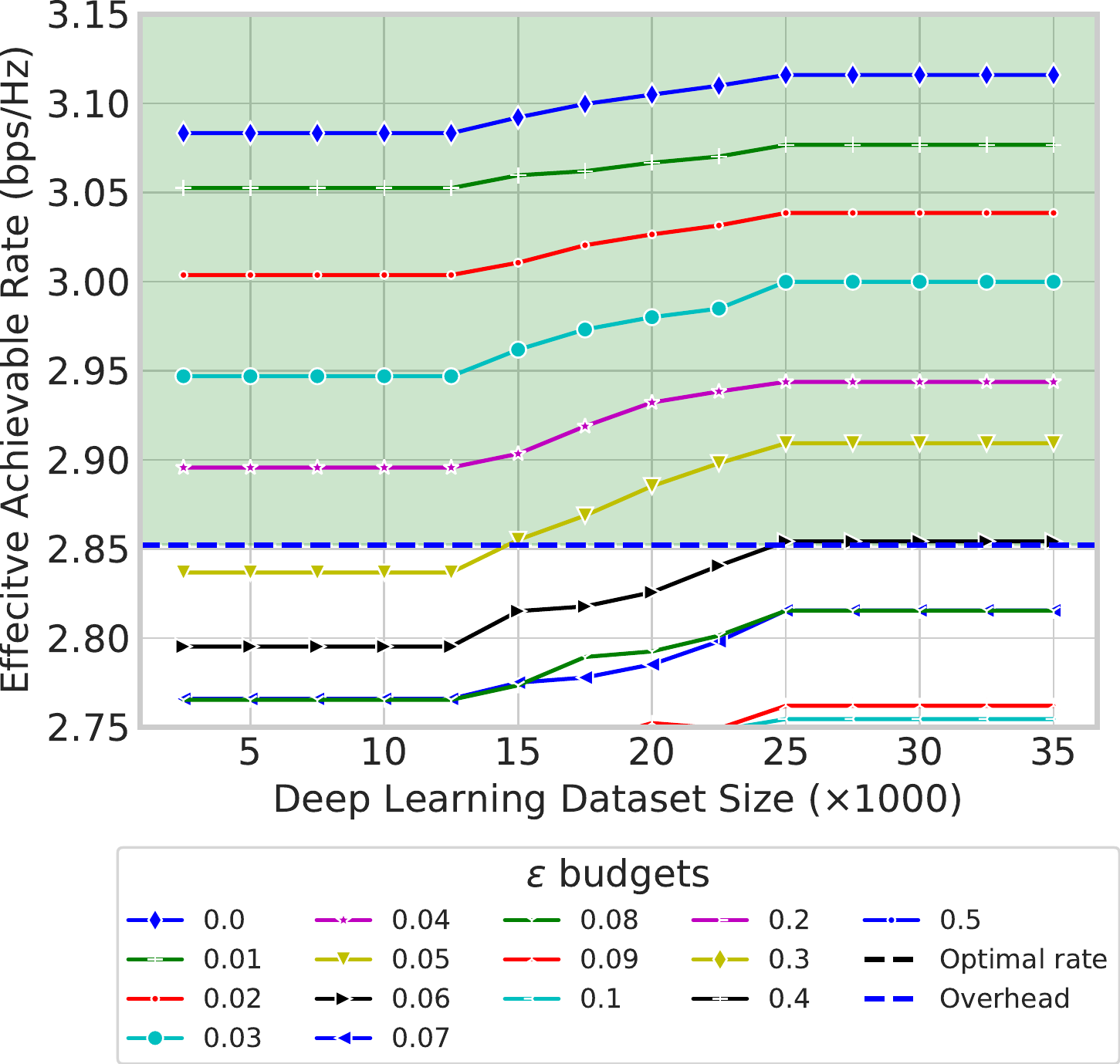}
         \caption{Undefended}
	    \label{fig:O1_undefended}
     \end{subfigure}
     \begin{subfigure}[b]{0.49\linewidth}
         \centering
         \captionsetup{justification=centering}
         \includegraphics[width=1\linewidth]{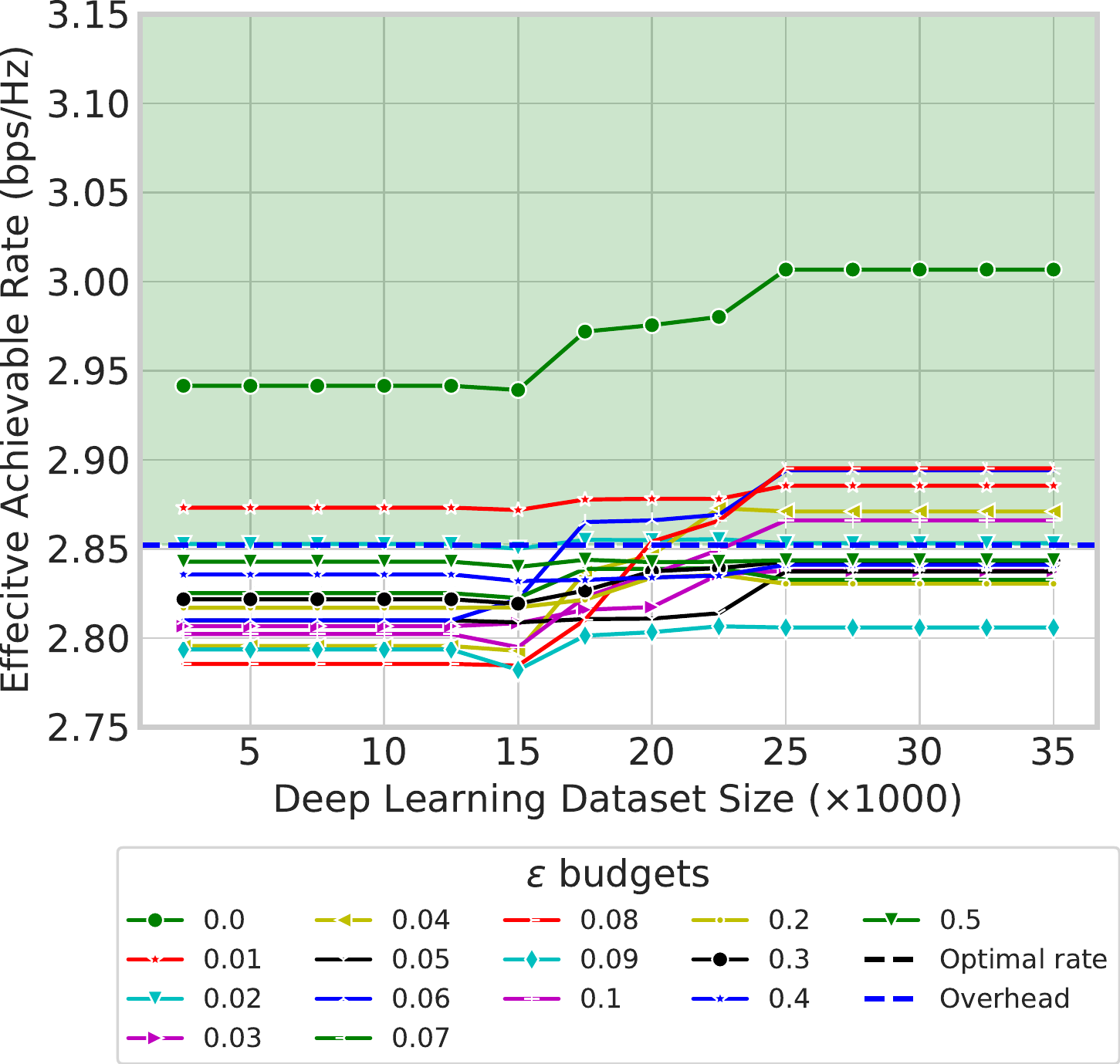}
         \caption{Defended}
	    \label{fig:O1_low_uncertainty}
     \end{subfigure}
	\caption{Beamforming codeword DL model results for Scenario-1 (O1) for different values of $ \epsilon $.}
	\label{fig:all_O1}
\end{figure}

\begin{figure}[!htbp]
    \centering
    \begin{subfigure}[b]{0.49\linewidth}
         \centering
         \captionsetup{justification=centering}
         \includegraphics[width=1\linewidth]{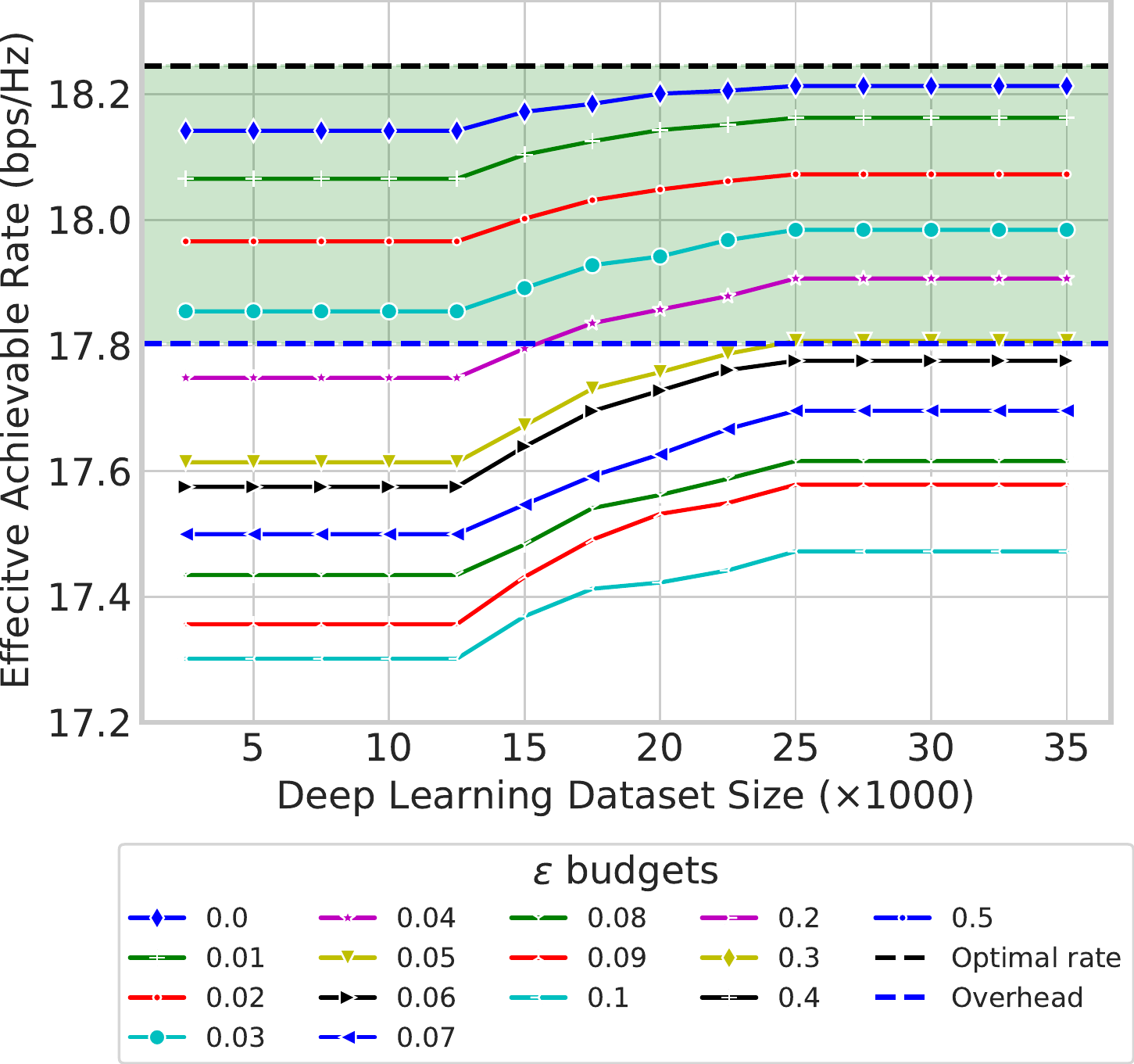}
         \caption{Undefended}
	    \label{fig:I1_2p5_undefended}
     \end{subfigure}
     \begin{subfigure}[b]{0.49\linewidth}
         \centering
         \captionsetup{justification=centering}
         \includegraphics[width=1\linewidth]{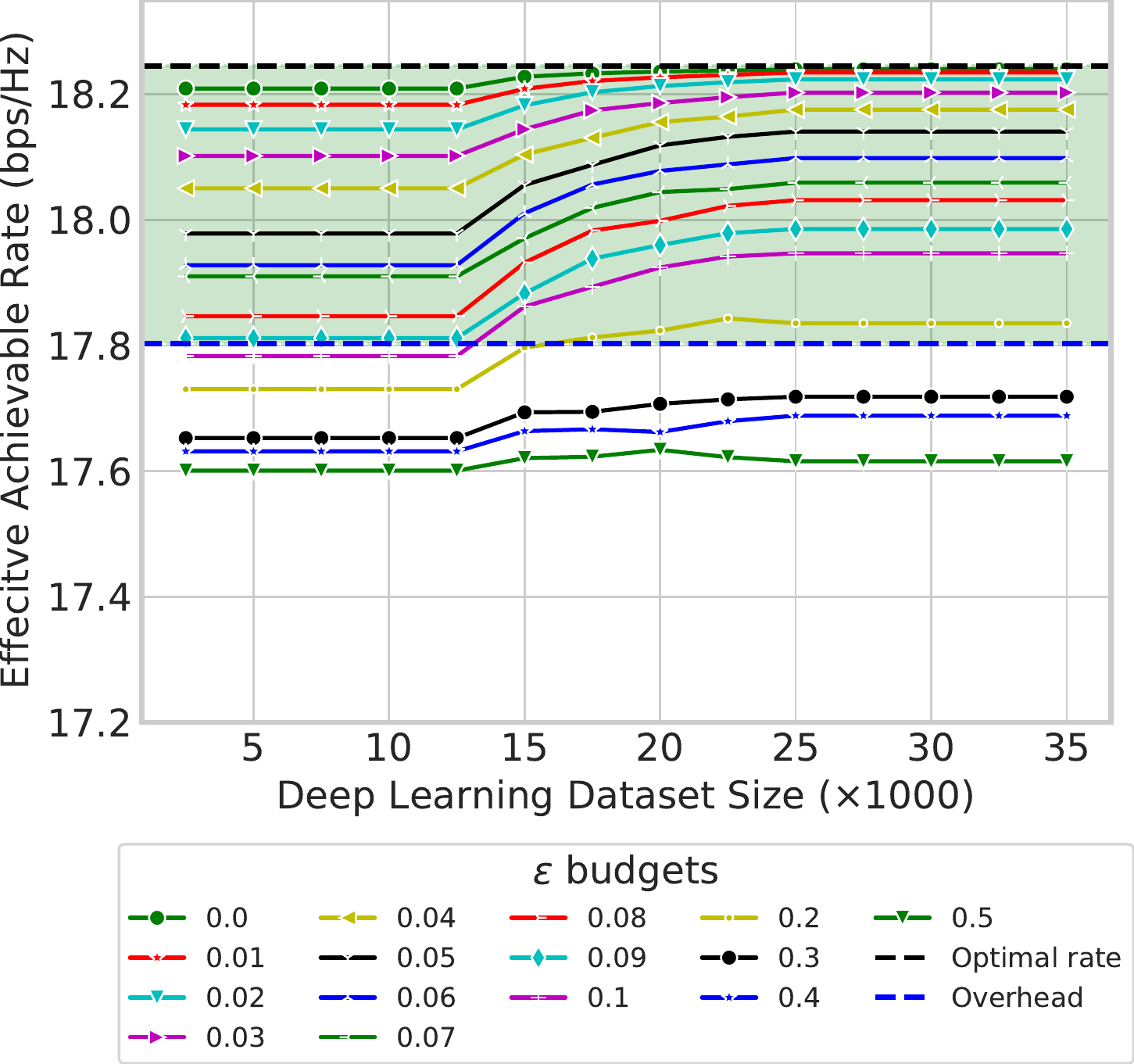}
         \caption{Defended}
	    \label{fig:I1_2p5_low_uncertainty}
     \end{subfigure}
	\caption{Beamforming codeword DL model results for Scenario-2 (I1\_2p5) for different values of $ \epsilon $.}
	\label{fig:all_I1_2p5}
\end{figure}

\begin{figure}[!htbp]
    \centering
    \begin{subfigure}[b]{0.49\linewidth}
         \centering
         \captionsetup{justification=centering}
         \includegraphics[width=1\linewidth]{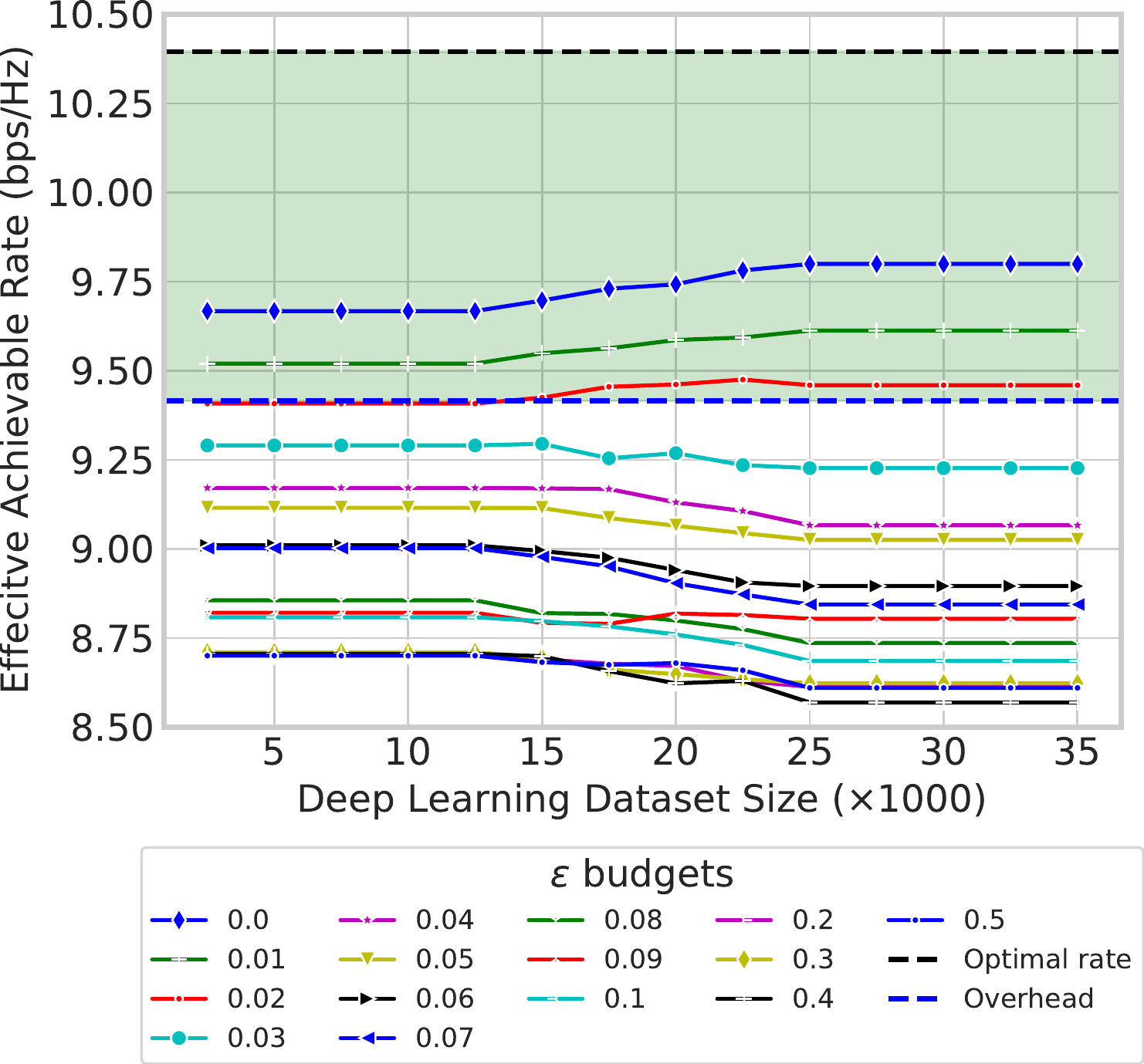}
         \caption{Undefended}
	    \label{fig:I3_60_undefended}
     \end{subfigure}
     \begin{subfigure}[b]{0.49\linewidth}
         \centering
         \captionsetup{justification=centering}
         \includegraphics[width=1\linewidth]{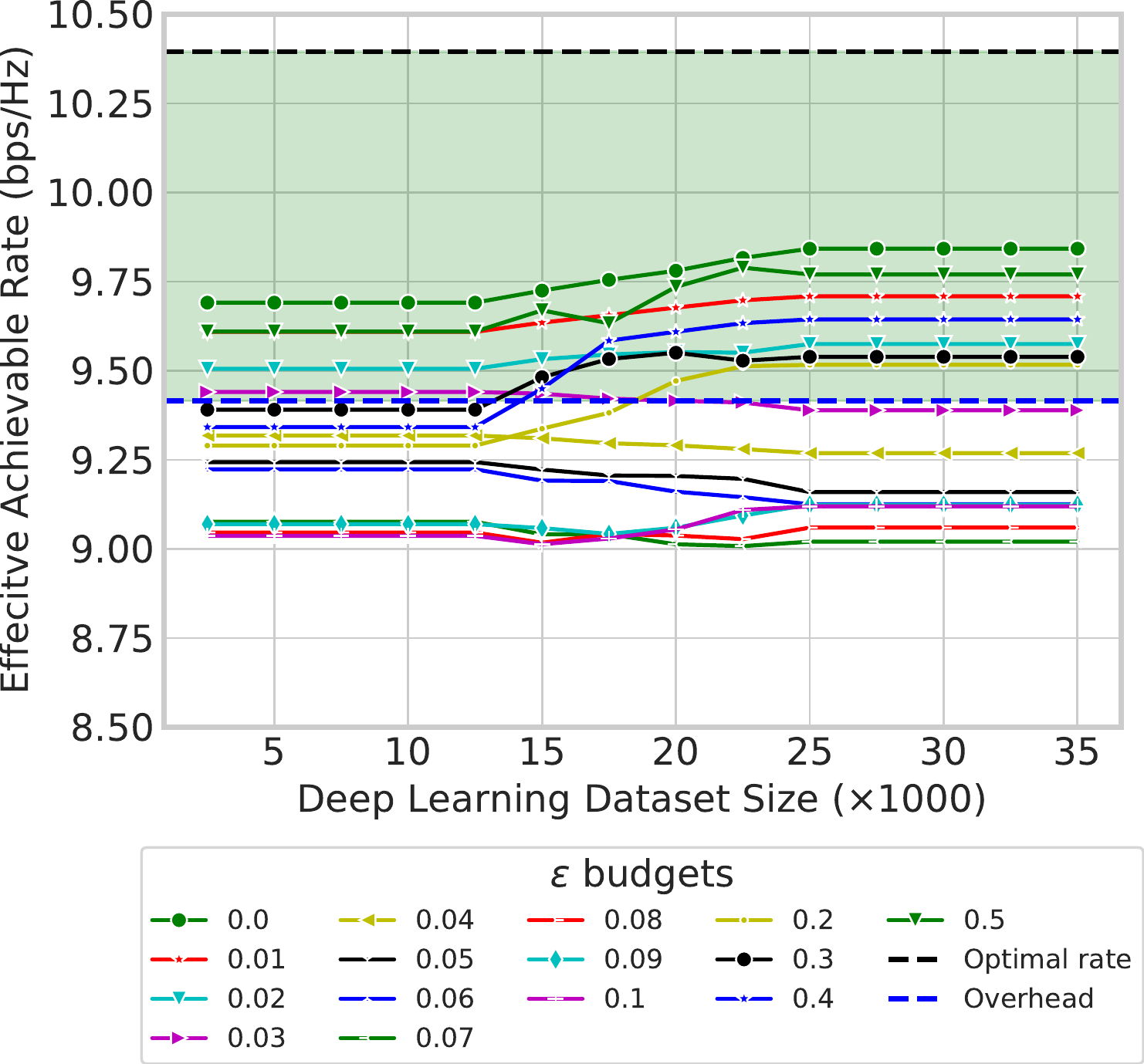}
         \caption{Defended}
	    \label{fig:I3_60_low_uncertainty}
     \end{subfigure}
	\caption{Beamforming codeword DL model results for Scenario-3 (I3\_60) for different values of $ \epsilon $.}
	\label{fig:I3_60}
\end{figure}

Figures \ref{fig:all_O1}-\ref{fig:I3_60} shows the original undefended and defended model under FGSM attack. Genie-aided coordinated beamforming is the optimal beamforming vector with no training overhead and baseline coordinated beamforming calculates with conventional communication system tools \cite{8395149}. According to the simulation results, the DL model's predictions are very close to the original value. We have used $l_{\infty}$ norm as the distance metric, which shows the maximum allowable perturbation amount for each item in the input vector $\mathbf{x}$. The green area in the figures shows the acceptable range between optimal and overhead limits. As can be seen from the figures, the predictive performance results of the vulnerable models fall below the green zone with shallow epsilon values. For the results of the models that have been robust with adversarial training to show low performance (i.e. to fall below the green zone), the attacker must use a very high epsilon value. A high epsilon value (i.e., more noise) will cause the attacker to be exposed. Therefore, we can say that the adversarial training method protects the DL model against the FGSM attack.

According to the results, the undefended RF beamforming codeword prediction model is vulnerable to the FGSM attack. The MSE performance result of the model under attack is approximately 40 (i.e. $\frac{0.00843 (Normal)}{0.00021 (Attacked)} \approx 40.14$) times higher.

\subsection{Results for RQ2}\label{sec:res_rq2}
\begin{figure}[htbp!]
    \centering
    \includegraphics[width=0.8\linewidth]{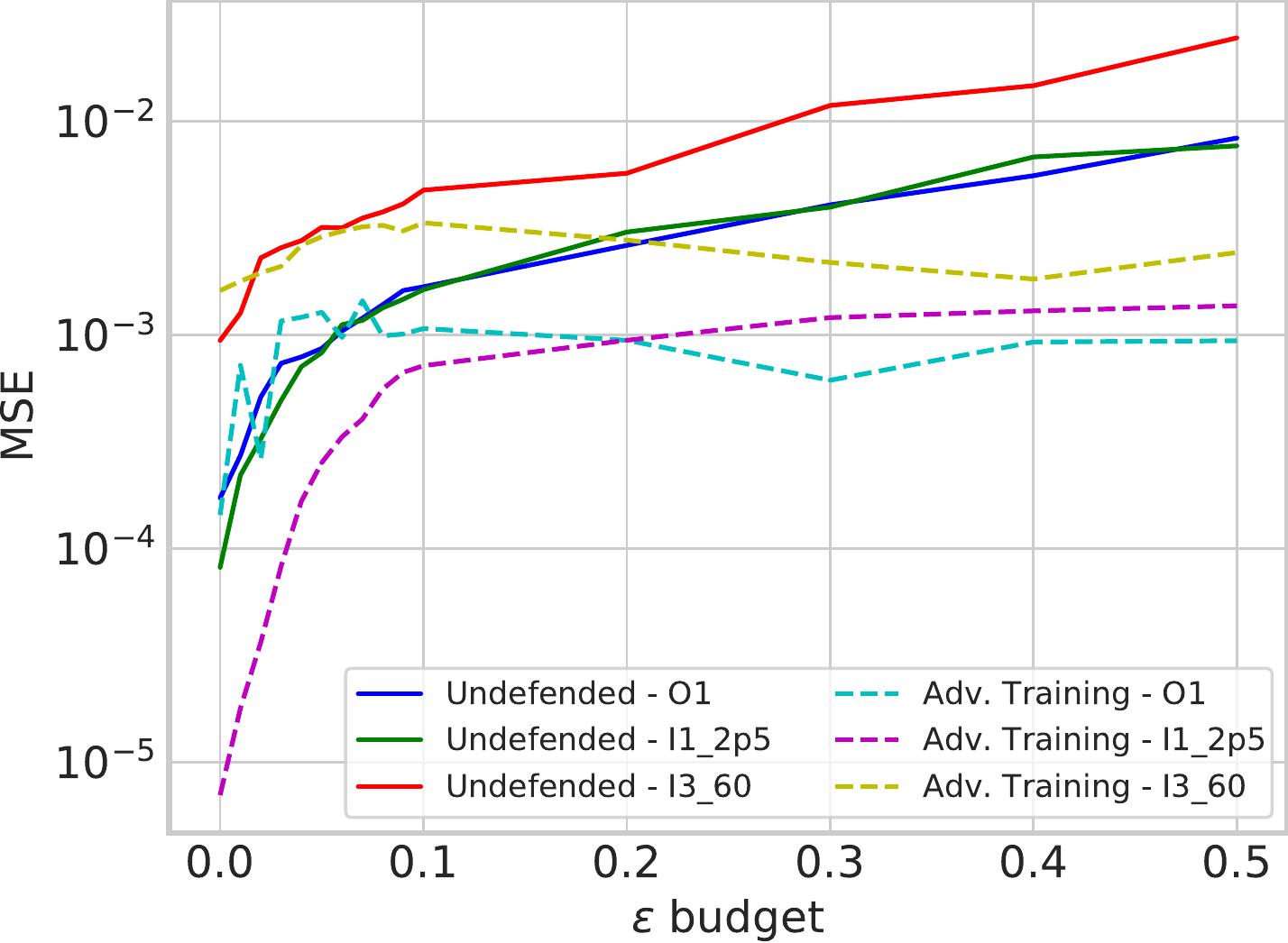}
    \caption{The performance results for all scenarios.}
    \label{fig:rq2}
\end{figure}

Adversarial training is a popularly advised defense mechanism \cite{2021arXiv210201356B,2014arXiv1412.6572G} that proposes generating adversarial instances using the victim model's loss function and then re-training the model with the newly generated adversarial instances and their respective outputs. This approach has proved to be effective in protecting DL models from adversarial machine learning attacks. Figure \ref{fig:rq2} shows the MSE of the performance results for all scenarios with the FGSM attack. According to the figure, defended (adversarial trained) model's MSE values becomes steady-state after a specific $\epsilon$ value. On the other hand, the undefended model's MSE values continue to increase.

Table \ref{tab:eps_res} shows the beamforming codeword prediction results for all scenarios for different values of $\epsilon$. The overhead (lower) values for each scenario are 2.86 for O1, 17.81 for I1\_2p5 and 9.41 for I3\_60. According to the table, the attacker can manipulate the DL model with $\epsilon=0.06$ for the O1 scenario. The undefended model's prediction result is 2.85426, which is lower than the overhead value that is 2.86. Similarly, the $\epsilon$ values for the successful attacks are 0.05 for I1\_2p5 and 0.03 for I3\_60.

\begin{table*}[!htbp]
\centering
\caption{Beamforming codeword prediction results for all scenarios for different values of $ \epsilon $.}
\label{tab:eps_res}
\begin{tabular}{|c|rr||rr||rr|}
\hline
& \multicolumn{2}{c||}{O1} & \multicolumn{2}{c||}{I1\_2p5} & \multicolumn{2}{c|}{I3\_60} \\
 $\epsilon$ &  \textbf{Undef.} & \textbf{Def.} &  \textbf{Undef.} & \textbf{Def.} &  \textbf{Undef.} & \textbf{Def.}  \\
\hline
0.00 & 3.11594 & 3.00674  & 18.21316 & 18.23962 & 9.79951 & 9.84233 \\
0.01 & 3.07667 & 2.88551  & 18.16247 & 18.23461 & 9.61264 & 9.70870 \\
0.02 & 3.03850 & 2.85323  & 18.07253 & 18.22359 & 9.45940 & 9.57484 \\
0.03 & 2.99991 & 2.83773  & 17.98412 & 18.20262 & 9.22683 & 9.38894 \\
0.04 & 2.94373 & 2.87103  & 17.90667 & 18.17553 & 9.06634 & 9.26890 \\
0.05 & 2.90929 & 2.83736  & 17.80715 & 18.14055 & 9.02542 & 9.15959 \\
0.06 & 2.85426 & 2.89408  & 17.77563 & 18.09840 & 8.89587 & 9.12542 \\
0.07 & 2.81551 & 2.83269  & 17.69611 & 18.05923 & 8.84408 & 9.02040 \\
0.08 & 2.81535 & 2.89524  & 17.61604 & 18.03128 & 8.73619 & 9.05983 \\
0.09 & 2.76199 & 2.80589  & 17.57838 & 17.98539 & 8.80412 & 9.12413 \\
0.10 & 2.75456 & 2.86601  & 17.47225 & 17.94672 & 8.68579 & 9.11946 \\
0.20 & 2.65541 & 2.83044  & 17.01810 & 17.83537 & 8.61330 & 9.51652 \\
0.30 & 2.58553 & 2.84275  & 16.61877 & 17.71839 & 8.62319 & 9.53898 \\
0.40 & 2.56208 & 2.84106  & 16.36651 & 17.68828 & 8.56926 & 9.64404 \\
0.50 & 2.57365 & 2.84363  & 16.26341 & 17.61559 & 8.61028 & 9.77017 \\
\hline
\end{tabular}
\end{table*}

\subsection{Threats to Validity}\label{sec:threats_to_val}
A key \textit{external validity} threat is related to the generalization of results \cite{a06430d231c644d8a1278057a2dd6956}. We used only the RF beamforming dataset in our experiments, and we need more case studies to generalize the results. Moreover, the dataset reflects different types of millimeter-wave beams. 

Our key \textit{construct validity} threat is related to the selection of attack type FGSM. Nevertheless, note that this attack is from the literature \cite{a06430d231c644d8a1278057a2dd6956} and applied to several DL usage domains. In the future, we will conduct dedicated empirical studies to investigate more adversarial machine learning attacks systematically.

Our main \textit{conclusion validity} threat is due to finding the best attack budget $\epsilon$ that is responsible for manipulating the legitimate user's signal for poisoning the beamforming prediction model. To mitigate this threat, we repeated each experiment 20 times to reduce the probability that the results were obtained by chance. In a standard neural network training, all weights are initialized uniformly at random. In the second stage, using optimization, these weights are updated to fit the classification problem. Since the training started with a probabilistic approach, there is a possibility of facing optimization's local minimum problem. We repeat the training 20 times to find the $\epsilon$ value that gives the best attack result to eliminate the local minimum problem. In each repetition, the weights were initialized uniformly at random but with different values. If the optimization function failed to find the global minimum in the next experiment, it is likely to see it as the weights have been initialized with different values.

\section{Conclusions and Future Works}\label{sec:conclusion}
This study emphasizes cyber-security issues related to RF beamforming codeword prediction models' vulnerabilities by satisfying the following research questions: 
(1) Is the DL-based RF beamforming codeword predictor vulnerable against adversarial addressees machine learning attacks? (2) Is the iterative adversarial training approach a mitigation method for the adversarial attacks in beamforming prediction? 
The experiments were performed with the DeepMIMO's \textit{O1, I1\_2p5 and I3\_60 ray-tracing scenarios} to answer these questions are performed. 
Our results confirm that the original model is vulnerable to a modified FGSM type of attack. The proposed mitigation method is the iterative adversarial training approach. Our empirical results also show that iterative adversarial training successfully increases the RF beamforming prediction performance and creates a more accurate predictor, suggesting that the strategy can improve the predictor's performance. The attacker must increase the epsilon value from 0.05 to 0.06 for the O1 scenario, from 0.04 to 0.2 for the I1\_2p5 scenario, and from 0.02 to 0.02 for the I3\_60 scenario in order to perform a successful attack. Due to the higher epsilon value, the probability of the attack being detected by another security component also increases. As future work, the outcomes of this study has potential to be further used and developed for other future studies to gain more insights in the field of 6G, where adversarial DL based cyber-security issues will be advanced.

\bibliographystyle{ieeetr}
\bibliography{myref}

\begin{thebibliography}{10}

\bibitem{6515173}
T.~S. Rappaport, S.~Sun, R.~Mayzus, H.~Zhao, Y.~Azar, K.~Wang, G.~N. Wong,
  J.~K. Schulz, M.~Samimi, and F.~Gutierrez, ``Millimeter wave mobile
  communications for 5{G} cellular: It will work!,'' {\em IEEE Access}, vol.~1,
  pp.~335--349, 2013.

\bibitem{9040431}
H.~{Viswanathan} and P.~E. {Mogensen}, ``Communications in the 6{G} era,'' {\em
  IEEE Access}, vol.~8, pp.~57063--57074, 2020.

\bibitem{ccatak2016waveform}
E.~Catak and L.~Durak-Ata, ``Waveform design considerations for 5{G} wireless
  networks,'' {\em Towards 5{G} Wireless Networks-A Physical Layer
  Perspective}, pp.~27--48, 2016.

\bibitem{CATAK2017184}
E.~Catak and L.~Durak-Ata, ``Adaptive filterbank-based multi-carrier waveform
  design for flexible data rates,'' {\em Computers \& Electrical Engineering},
  vol.~61, pp.~184--194, 2017.

\bibitem{6736750}
W.~{Roh}, J.~{Seol}, J.~{Park}, B.~{Lee}, J.~{Lee}, Y.~{Kim}, J.~{Cho},
  K.~{Cheun}, and F.~{Aryanfar}, ``Millimeter-wave beamforming as an enabling
  technology for 5{G} cellular communications: theoretical feasibility and
  prototype results,'' {\em IEEE Communications Magazine}, vol.~52, no.~2,
  pp.~106--113, 2014.

\bibitem{6815892}
V.~Jungnickel, K.~Manolakis, W.~Zirwas, B.~Panzner, V.~Braun, M.~Lossow,
  M.~Sternad, R.~Apelfröjd, and T.~Svensson, ``The role of small cells,
  coordinated multipoint, and massive mimo in 5g,'' {\em IEEE Communications
  Magazine}, vol.~52, no.~5, pp.~44--51, 2014.

\bibitem{6848613}
F.~W. Vook, A.~Ghosh, and T.~A. Thomas, ``Mimo and beamforming solutions for 5g
  technology,'' in {\em 2014 IEEE MTT-S International Microwave Symposium
  (IMS2014)}, pp.~1--4, 2014.

\bibitem{6736746}
F.~Boccardi, R.~W. Heath, A.~Lozano, T.~L. Marzetta, and P.~Popovski, ``Five
  disruptive technology directions for 5{G},'' {\em IEEE Communications
  Magazine}, vol.~52, no.~2, pp.~74--80, 2014.

\bibitem{8922617}
T.~Huang, W.~Yang, J.~Wu, J.~Ma, X.~Zhang, and D.~Zhang, ``A survey on green 6g
  network: Architecture and technologies,'' {\em IEEE Access}, vol.~7,
  pp.~175758--175768, 2019.

\bibitem{8395149}
A.~{Alkhateeb}, S.~{Alex}, P.~{Varkey}, Y.~{Li}, Q.~{Qu}, and D.~{Tujkovic},
  ``Deep learning coordinated beamforming for highly-mobile millimeter wave
  systems,'' {\em IEEE Access}, vol.~6, pp.~37328--37348, 2018.

\bibitem{9034044}
M.~S. {Sim}, Y.~{Lim}, S.~H. {Park}, L.~{Dai}, and C.~{Chae}, ``Deep
  learning-based mmwave beam selection for 5{G NR}7/6{G} with sub-6 {GH}z
  channel information: Algorithms and prototype validation,'' {\em IEEE
  Access}, vol.~8, pp.~51634--51646, 2020.

\bibitem{8644288}
Y.~{Wang}, A.~{Klautau}, M.~{Ribero}, M.~{Narasimha}, and R.~W. {Heath},
  ``Mmwave vehicular beam training with situational awareness by machine
  learning,'' in {\em 2018 IEEE Globecom Workshops (GC Wkshps)}, pp.~1--6,
  2018.

\bibitem{8766143}
Z.~Zhang, Y.~Xiao, Z.~Ma, M.~Xiao, Z.~Ding, X.~Lei, G.~K. Karagiannidis, and
  P.~Fan, ``6{G} wireless networks: Vision, requirements, architecture, and key
  technologies,'' {\em IEEE Vehicular Technology Magazine}, vol.~14, no.~3,
  pp.~28--41, 2019.

\bibitem{8808168}
K.~B. Letaief, W.~Chen, Y.~Shi, J.~Zhang, and Y.-J.~A. Zhang, ``The roadmap to
  6g: Ai empowered wireless networks,'' {\em IEEE Communications Magazine},
  vol.~57, no.~8, pp.~84--90, 2019.

\bibitem{9240722}
C.~{Yizhan}, W.~{Zhong}, H.~{Da}, and L.~{Ruosen}, ``6{G} is coming :
  Discussion on key candidate technologies and application scenarios,'' in {\em
  2020 International Conference on Computer Communication and Network Security
  (CCNS)}, pp.~59--62, 2020.

\bibitem{lu2020security}
Y.~Lu, ``Security in 6{G}: The prospects and the relevant technologies,'' {\em
  Journal of Industrial Integration and Management}, vol.~5, no.~03,
  pp.~271--289, 2020.

\bibitem{wang2020security}
M.~Wang, T.~Zhu, T.~Zhang, J.~Zhang, S.~Yu, and W.~Zhou, ``Security and privacy
  in 6{G} networks: New areas and new challenges,'' {\em Digital Communications
  and Networks}, vol.~6, no.~3, pp.~281--291, 2020.

\bibitem{chorti2021context}
A.~Chorti, A.~N. Barreto, S.~Kopsell, M.~Zoli, M.~Chafii, P.~Sehier,
  G.~Fettweis, and H.~V. Poor, ``Context-aware security for 6{G} wireless the
  role of physical layer security,'' {\em arXiv preprint arXiv:2101.01536},
  2021.

\bibitem{9311932}
Y.~{Xiao}, G.~{Shi}, Y.~{Li}, W.~{Saad}, and H.~V. {Poor}, ``Toward
  self-learning edge intelligence in 6{G},'' {\em IEEE Communications
  Magazine}, vol.~58, no.~12, pp.~34--40, 2020.

\bibitem{kuzlu2021role}
M.~Kuzlu, C.~Fair, and O.~Guler, ``Role of artificial intelligence in the
  internet of things ({IoT}) cybersecurity,'' {\em Discover Internet of
  Things}, vol.~1, no.~1, pp.~1--14, 2021.

\bibitem{9146540}
Y.~{Sun}, J.~{Liu}, J.~{Wang}, Y.~{Cao}, and N.~{Kato}, ``When machine learning
  meets privacy in 6{G}:{A} survey,'' {\em IEEE Communications Surveys
  Tutorials}, vol.~22, no.~4, pp.~2694--2724, 2020.

\bibitem{9143575}
Q.~{Liu}, J.~{Guo}, C.~K. {Wen}, and S.~{Jin}, ``Adversarial attack on
  {DL}-based massive {MIMO CSI} feedback,'' {\em Journal of Communications and
  Networks}, vol.~22, no.~3, pp.~230--235, 2020.

\bibitem{6002149}
Z.~Yang, Y.~Yue, Y.~Yang, Y.~Peng, X.~Wang, and W.~Liu, ``Study and application
  on the architecture and key technologies for iot,'' in {\em 2011
  International Conference on Multimedia Technology}, pp.~747--751, 2011.

\bibitem{6424332}
R.~Khan, S.~U. Khan, R.~Zaheer, and S.~Khan, ``Future internet: The internet of
  things architecture, possible applications and key challenges,'' in {\em 2012
  10th International Conference on Frontiers of Information Technology},
  pp.~257--260, 2012.

\bibitem{7123563}
A.~Al-Fuqaha, M.~Guizani, M.~Mohammadi, M.~Aledhari, and M.~Ayyash, ``Internet
  of things: A survey on enabling technologies, protocols, and applications,''
  {\em IEEE Communications Surveys Tutorials}, vol.~17, no.~4, pp.~2347--2376,
  2015.

\bibitem{RAKOTONIRAINY201678}
A.~Rakotonirainy, O.~Orfila, and D.~Gruyer, ``Reducing driver's behavioural
  uncertainties using an interdisciplinary approach: Convergence of quantified
  self, automated vehicles, internet of things and artificial intelligence.,''
  {\em IFAC-PapersOnLine}, vol.~49, no.~32, pp.~78--82, 2016.
\newblock Cyber-Physical \& Human-Systems CPHS 2016.

\bibitem{rahman2018enabling}
H.~Rahman and R.~Rahmani, ``Enabling distributed intelligence assisted future
  internet of things controller (fitc),'' {\em Applied computing and
  informatics}, vol.~14, no.~1, pp.~73--87, 2018.

\bibitem{vermesan2017internet}
O.~Vermesan, A.~Br{\"o}ring, E.~Tragos, M.~Serrano, D.~Bacciu, S.~Chessa,
  C.~Gallicchio, A.~Micheli, M.~Dragone, A.~Saffiotti, {\em et~al.}, ``Internet
  of robotic things: converging sensing/actuating, hypoconnectivity, artificial
  intelligence and iot platforms,'' 2017.

\bibitem{7971869}
C.~Kolias, G.~Kambourakis, A.~Stavrou, and J.~Voas, ``Ddos in the iot: Mirai
  and other botnets,'' {\em Computer}, vol.~50, no.~7, pp.~80--84, 2017.

\bibitem{AMMAR20188}
M.~Ammar, G.~Russello, and B.~Crispo, ``Internet of things: A survey on the
  security of iot frameworks,'' {\em Journal of Information Security and
  Applications}, vol.~38, pp.~8--27, 2018.

\bibitem{8378034}
S.~Surendran, A.~Nassef, and B.~D. Beheshti, ``A survey of cryptographic
  algorithms for iot devices,'' in {\em 2018 IEEE Long Island Systems,
  Applications and Technology Conference (LISAT)}, pp.~1--8, 2018.

\bibitem{8319238}
S.~Alharbi, P.~Rodriguez, R.~Maharaja, P.~Iyer, N.~Bose, and Z.~Ye, ``Focus: A
  fog computing-based security system for the internet of things,'' in {\em
  2018 15th IEEE Annual Consumer Communications Networking Conference (CCNC)},
  pp.~1--5, 2018.

\bibitem{MAHDAVINEJAD2018161}
M.~S. Mahdavinejad, M.~Rezvan, M.~Barekatain, P.~Adibi, P.~Barnaghi, and A.~P.
  Sheth, ``Machine learning for internet of things data analysis: a survey,''
  {\em Digital Communications and Networks}, vol.~4, no.~3, pp.~161--175, 2018.

\bibitem{shanbhogue2017survey}
R.~Shanbhogue and B.~Beena, ``Survey of data mining (dm) and machine learning
  (ml) methods on cyber security,'' {\em Indian Journal of Science and
  Technology}, vol.~10, no.~35, pp.~1--7, 2017.

\bibitem{saljoughi2017attacks}
A.~S. Saljoughi, M.~Mehrvarz, and H.~Mirvaziri, ``Attacks and intrusion
  detection in cloud computing using neural networks and particle swarm
  optimization algorithms,'' {\em Emerging Science Journal}, vol.~1, no.~4,
  pp.~179--191, 2017.

\bibitem{1447503}
H.~Urkowitz, ``Energy detection of unknown deterministic signals,'' {\em
  Proceedings of the IEEE}, vol.~55, no.~4, pp.~523--531, 1967.

\bibitem{2016arXiv161101236K}
A.~{Kurakin}, I.~{Goodfellow}, and S.~{Bengio}, ``{Adversarial Machine Learning
  at Scale},'' {\em arXiv e-prints}, p.~arXiv:1611.01236, Nov. 2016.

\bibitem{8965459}
M.~{Aladag}, F.~O. {Catak}, and E.~{Gul}, ``Preventing data poisoning attacks
  by using generative models,'' in {\em 2019 1st International Informatics and
  Software Engineering Conference (UBMYK)}, pp.~1--5, 2019.

\bibitem{2021arXiv210204150F}
O.~{Faruk Tuna}, F.~{Ozgur Catak}, and M.~{Taner Eskil}, ``{Exploiting
  epistemic uncertainty of the deep learning models to generate adversarial
  samples},'' {\em arXiv e-prints}, p.~arXiv:2102.04150, Feb. 2021.

\bibitem{2018arXiv180902861D}
A.~{Demontis}, M.~{Melis}, M.~{Pintor}, M.~{Jagielski}, B.~{Biggio},
  A.~{Oprea}, C.~{Nita-Rotaru}, and F.~{Roli}, ``{Why Do Adversarial Attacks
  Transfer? Explaining Transferability of Evasion and Poisoning Attacks},''
  {\em arXiv e-prints}, p.~arXiv:1809.02861, Sept. 2018.

\bibitem{6979963}
A.~{Alkhateeb}, J.~{Mo}, N.~{Gonzalez-Prelcic}, and R.~W. {Heath}, ``{MIMO}
  precoding and combining solutions for millimeter-wave systems,'' {\em IEEE
  Communications Magazine}, vol.~52, no.~12, pp.~122--131, 2014.

\bibitem{7742901}
V.~{Va}, J.~{Choi}, and R.~W. {Heath}, ``The impact of beamwidth on temporal
  channel variation in vehicular channels and its implications,'' {\em IEEE
  Transactions on Vehicular Technology}, vol.~66, no.~6, pp.~5014--5029, 2017.

\bibitem{deepmimo_adam}
``Deepmimo ray tracing scenarios,'' 2021.

\bibitem{alkhateeb2019deepmimo}
A.~Alkhateeb, ``Deep{MIMO}: A generic deep learning dataset for millimeter wave
  and massive {MIMO} applications,'' 2019.

\bibitem{2021arXiv210201356B}
T.~{Bai}, J.~{Luo}, J.~{Zhao}, B.~{Wen}, and Q.~{Wang}, ``{Recent Advances in
  Adversarial Training for Adversarial Robustness},'' {\em arXiv e-prints},
  p.~arXiv:2102.01356, Feb. 2021.

\bibitem{2014arXiv1412.6572G}
I.~J. {Goodfellow}, J.~{Shlens}, and C.~{Szegedy}, ``{Explaining and Harnessing
  Adversarial Examples},'' {\em arXiv e-prints}, p.~arXiv:1412.6572, Dec. 2014.

\bibitem{a06430d231c644d8a1278057a2dd6956}
P.~Runeson, M.~H{\"o}st, R.~Austen, and B.~Regnell, {\em Case Study Research in
  Software Engineering – Guidelines and Examples}.
\newblock United States: John Wiley and Sons Inc., 2012.

\end{thebibliography}

\end{document}